\documentclass[graybox]{svmult}

\usepackage{mathptmx}       
\usepackage{helvet}         
\usepackage{courier}        
\usepackage{type1cm}        
\usepackage{makeidx}         
\usepackage{graphicx}        
\usepackage{multicol}        
\usepackage[bottom]{footmisc}

\makeindex             

\newcommand{\zi}[1]{\textit{#1}}
\newcommand{\xb}[1]{\textbf{#1}}
\newcommand{\xf}[1]{\textsf{#1}}

\newcommand{\xt}[1]{\texttt{#1}}

\newcommand{\req}[1]{\xt{#1}}
\newcommand{\REfull}{Requirements Engineering}
\newcommand{\RE}{RE}
\newcommand{\RPfull}{Requirements Problem}
\newcommand{\RP}{RP}
\newcommand{\ASfull}{Adaptive System}
\newcommand{\AS}{AS}
\newcommand{\RPASfull}{\RPfull{} for \ASfull}
\newcommand{\RPAS}{RPAS}
\newcommand{\ZJ}{ZJ}
\newcommand{\ZJRPfull}{Default \RPfull}
\newcommand{\ZJRP}{DRP}
\newcommand{\SatisfactionCondition}{Satisfaction Condition}
\newcommand{\ConsistencyCondition}{Consistency Condition}
\newcommand{\OptimalityCondition}{Optimality Condition}
\newcommand{\nmdash}{\mid\negthinspace\sim}
\newcommand{\SolutionSpace}{Solution Space}
\newcommand{\ProblemSpace}{Problem Space}
\newcommand{\Specification}{Specification}
\newcommand{\Solution}{Solution}
\newcommand{\Criterion}{Criterion}
\newcommand{\Criteria}{Criteria}
\newcommand{\ProblemInstance}{Problem Instance}
\newcommand{\Parameter}{Parameter}
\newcommand{\Solve}{Solve}
\newcommand{\Depend}{Depend}

\newcommand{\OptimalSpecification}{Optimal Specification}
\newcommand{\DecisionRule}{Decision Rule}
\newcommand{\ROPfull}{Requirements Optimisation Problem}
\newcommand{\ROP}{ROP}
\newcommand{\DecisionSet}{Decision Set}
\newcommand{\RDPfull}{Requirements Design Problem}
\newcommand{\RDP}{RDP}
\newcommand{\ZJROPfull}{Revised Default \RPfull}
\newcommand{\ZJROP}{RDRP}
\newcommand{\MonitoringScope}{Monitoring Scope}
\newcommand{\ChangeScope}{Change Scope}
\newcommand{\MonitoredVariable}{Monitored Variable}

\newcommand{\RDPASfull}{\RDPfull{} for \ASfull}
\newcommand{\RDPAS}{RDPAS}
\newcommand{\ROPASfull}{\ROPfull{} for \ASfull}
\newcommand{\ROPAS}{ROPAS}
\newcommand{\EvolutionRequirement}{Evolution Requirement}
\newcommand{\PrR}{\xf{R}}
\newcommand{\PrK}{\xf{K}}
\newcommand{\PrS}{\xf{S}}
\newcommand{\PrKSR}{\xf{KSR}}
\newcommand{\PrSatisfaction}{\xf{Satisfaction}}
\newcommand{\PrConsistency}{\xf{Consistency}}
\newcommand{\PrDR}{\xf{Decision Rule}}
\newcommand{\PrOpt}{\xf{Optimality}}
\newcommand{\PropR}{\PrR{} Property}
\newcommand{\PropK}{\PrK{} Property}
\newcommand{\PropS}{\PrS{} Property}
\newcommand{\PropKSR}{\PrKSR{} Property}
\newcommand{\PropSatisfaction}{\PrSatisfaction{} Property}
\newcommand{\PropConsistency}{\PrConsistency{} Property}
\newcommand{\PropDR}{\PrDR{} Property}
\newcommand{\PropOpt}{\xf{Optimality Property}}
\newcommand{\CLabel}[2]{\xb{[}#1\xb{: #2]}}
\begin{document}
\title*{\RPfull{} and Solution Concepts for \ASfull s Engineering, and their Relationship to Mathematical Optimisation, Decision Analysis, and Expected Utility Theory}
\titlerunning{\RPfull{} and Solution Concepts for \ASfull s Engineering}
\author{Ivan J. Jureta}
\authorrunning{I. J. Jureta}
\institute{
Ivan J. Jureta \at Fonds de la Recherche Scientifique -- FNRS and Department of Business Administration, University of Namur, \email{ivan.jureta@unamur.be}
}
\maketitle
\abstract*{
\REfull{} (\RE) focuses on eliciting, modelling, and analyzing the requirements and environment of a system-to-be in order to design its specification. The design of the specification, usually called the \RPfull{} (\RP), is a complex problem solving task, as it involves, for each new system-to-be, the discovery and exploration of, and decision making in, new and ill-defined problem and solution spaces. The default \RP{} in \RE{} is to design a specification of the system-to-be which (i) is consistent with given requirements and  conditions of its environment, and (ii) together with environment conditions satisfies requirements. This paper (i) shows that the \RPASfull{} (\RPAS) is different from, and is not a subclass of the default \RP, (ii) gives a formal definition of the \RPAS, and (iii) discusses implications for future research.}
\abstract{
\REfull{} (\RE) focuses on eliciting, modelling, and analyzing the requirements and environment of a system-to-be in order to design its specification. The design of the specification, usually called the \RPfull{} (\RP), is a complex problem solving task, as it involves, for each new system-to-be, the discovery and exploration of, and decision making in, new and ill-defined problem and solution spaces. The default \RP{} in \RE{} is to design a specification of the system-to-be which (i) is consistent with given requirements and  conditions of its environment, and (ii) together with environment conditions satisfies requirements. This paper (i) shows that the \RPASfull{} (\RPAS) is different from, and is not a subclass of the default \RP, (ii) gives a formal definition of the \RPAS, and (iii) discusses implications for future research.}

\section{Introduction}\label{s:introduction}

\subsection{Domain: \REfull}\label{s:introduction:domain}
\xb{\REfull{} (\RE)} focuses on eliciting, modelling, and analysing the requirements and environment of a system-to-be in order to design its specification. 

It is on the basis of its specification that the system is built, updated, changed, its new releases planned, made, announced, rolled out. Specifications can take different forms, ranging from minimalistic to-do lists that hint at expectations and subsume implicit engineering solutions, to elaborately structured documentation on contracts with employees and suppliers, responsibilities of positions in the value chain, guidelines for employee coordination and collaboration, as well as formal software specifications made for use with a model checker. 

The design of the specification, usually called the \xb{\RPfull{} (\RP)}, is a complex problem solving task, as it involves, for each new system-to-be, the discovery and exploration of, and decision making in, new and ill-defined problem and solution spaces. 

Difficulties involved in solving an \RP{} instance are illustrated by the variety of topics studied in \RE{} research, such as requirements elicitation \cite{goguen1993techniques, hickey2004unified, davis2006effectiveness}, categorization \cite{dardenne1993goal, zave1997four, jureta2008revisiting}, vagueness and ambiguity \cite{mylopoulos1992representing, letier2004reasoning, jureta2007clarifying}, prioritization \cite{karlsson1998evaluation, berander2005requirements, herrmann2008requirements}, negotiation \cite{leite1991requirements, boehm1995software, jureta2009analysis}, responsibility allocation \cite{dardenne1993goal, castro2002towards, fuxman2004specifying}, cost estimation \cite{boehm1984software, boehm2000software, sindre2005eliciting}, conflicts and inconsistency \cite{nuseibeh1994framework, heitmeyer1996automated, van1998managing}, comparison \cite{mylopoulos1992representing, letier2004reasoning, liaskos2010integrating}, satisfaction evaluation \cite{boehm1976quantitative, mylopoulos1992representing, krogstie1995towards}, operationalization \cite{giorgini2003reasoning, fuxman2004specifying, ernst2013agile}, traceability \cite{gotel1994analysis, ramesh2001toward, cleland2005utilizing}, and change \cite{cheng2009software, whittle2009relax, brun2009engineering}.

\RE{} issues are present when designing new and changing existing systems; they are there whatever the system class and domain, and regardless of the extent to which people are involved in the system: from autonomic Internet-scale clouds, to traditional desktop applications, industrial expert systems, and embedded software, all enabling anything from massive mobile apps ecosystems, global supply chains, medical processes, business processes, mobile gaming, and so on. Moreover, RE issues are present regardless of how the software in the system is designed and made, from a military waterfall to a startup's own agile dialect, and from organisations where developers talk directly to customers, to those where product designers, salespeople, or others mediate between requirements and code.

\subsection{Context: \ZJRPfull}\label{s:introduction:context}
The \zi{de facto} default view in \RE, is that the specification is produced \zi{incrementally}, starting from a limited set of incomplete, inconsistent, and imprecise information about the requirements and the system's operating environment, and that each design step reduces incompleteness, removes inconsistencies, and improves precision, towards the specification of the system \cite{boehm1988spiral, dardenne1993goal, greenspan1994formal, nuseibeh1994framework, finkelstein1994inconsistency, zave1997four, van2001goal, castro2002towards, robinson2003requirements, jureta2010techne, ernst2013agile}.

This important and general conceptualisation of the aim in \RE{} is most clearly formulated in Zave \& Jackson's seminal paper, ``Four dark corners of requirements engineering'' \cite{zave1997four}. Their view, denoted \ZJ{} hereafter, is echoed in some of the most influential research in the field, which both preceded and followed the said paper, including, for example, contributions from Boehm et al. \cite{boehm1988spiral,boehm1995software}, van Lamsweerde et al. \cite{dardenne1993goal, darimont1996formal, van1998managing, van2000handling, van2001goal, letier2004reasoning}, Mylopoulos et al. \cite{mylopoulos1992representing, greenspan1994formal, castro2002towards}, Robinson et al. \cite{robinson2003requirements}, Nuseibeh et al. \cite{nuseibeh1994framework, hunter1998managing}, to name some.

According to the \ZJ{} view, in any concrete systems engineering project, \RE{} is successfully completed when the following conditions are satisfied \cite{zave1997four}:

\begin{quote}
\begin{enumerate}
    \item{``There is a set $R$ of requirements. Each member of $R$ has been validated (checked informally) as acceptable to the customer, and $R$ as a whole has been validated as expressing all the customer's desires with respect to the software development project.}
    \item{There is a set $K$ of statements of domain knowledge. Each member of $K$ has been validated (checked informally) as true of the environment.}
    \item{There is a set SS of specifications. The members of $S$ do not constrain the environment; they are not stated in terms of any unshared actions or state components; and they do not refer to the future.}
    \item{A proof shows that $K, S \vdash R$. This proof ensures that an implementation of $S$ will satisfy the requirements.}
    \item{There is a proof that $S$ and $K$ are consistent. This ensures that the specification is internally consistent and consistent with the environment. Note that the two proofs together imply that $S$, $K$, and $R$ are consistent with each other.''}
\end{enumerate}
\end{quote}

If the satisfaction of these conditions marks the end of \RE{} in any systems engineering project, then we can give a compact formulation of the default problem that \RE{} should solve, which we call the \ZJRPfull{} hereafter:

\begin{definition}\label{d:zjrp}
\xb{\ZJRPfull{} (\ZJRP):} \xb{Given} a set $R$ of requirements, and a set $K$ of domain knowledge, \xb{find} a specification $S$, such that $S$ satisfies the following conditions:
\begin{enumerate}
    \item{There is a proof of $R$ from $K$ and $S$, written $K, S \vdash R$,}
    \item{$K$ and $S$ are consistent, written $K, S \not\vdash \bot$.}
\end{enumerate}
\end{definition}

\subsection{Issue: What if the System is an \ASfull?}\label{s:introduction:issue}
A system is an \xb{\ASfull{} (\AS)} if it can detect differences between its design-time and run-time requirements and environment conditions, uses feedback mechanisms to analyse these changes and decide, with or without human input, how to adjust its behaviour as a result. 

There is nothing in the \ZJRP{} which makes it specific to a system class. This would suggest that the \RP{} for \AS{} is a subclass of \ZJRP, in the sense that it is the \ZJRP, with some additional properties that make it specific to the \ASfull s class.

It is important to know whether \RPAS{} is a subclass of the \ZJRP{}. According to Zave \& Jackson, \ZJRP{} ``establish[es] minimum standards for what information should be represented in a requirements language'' \cite{zave1997four}. 

But this is not only important because of the interest in the design of languages for the representation of requirements, domain knowledge, and specifications of \AS s. 

More generally, if it is the valid conclusion, then there are existing \RE{} tools (representation languages, methods, algorithms, etc.) that should be used in \RE{} for \AS, and the open question is how to specialise them to \AS. 

If it is not the valid conclusion, then the issue is to know which of the existing research is relevant for solving \RPAS, both in \RE{} and elsewhere, and what new research is needed. In both cases, discussing the validity of the conclusion above should provide relevant input for future research on \RE{} for \AS, and relate it to the default view of \RE.

\subsection{Contributions: \RPASfull s and its Relationship to the \ZJRPfull}\label{s:introduction:contributions}
This paper has three parts:
\begin{enumerate}
    \item{Part one runs from Section \ref{s:minimality} to Section \ref{s:optimality-criteria-importance}. It argues that the \xb{\RPASfull{} (\RPAS)} is \zi{different} from the \ZJRP, and that it is \zi{not a subclass of} the \ZJRP. This is argued in the following steps:
    \begin{enumerate}
        \item{The starting point, developed in Section \ref{s:minimality}, is the observation that \ZJRP{} is the \zi{minimal} \RP, in the sense that if something is removed from it, there is no meaningful problem left to solve.}
        \item{Minimality suggests that there may be similarity between the \ZJRP{} and every other \RP, including the \RPAS. The second step of the argument, developed in Section \ref{s:uniqueness}, is the observation that the \ZJRP{} is not the superclass of all \RP s, despite its minimality. This is argued by showing that, if we want to compare specifications when solving an \RP, and we do it in order to choose the one that is somehow \zi{the optimal one}, then that \RP{} is not a subclass of the \ZJRP.}
        \item{The third step of the argument, in Section \ref{s:non-functional-requirements}, shows that to want the optimal specification as a solution to an \RP, is not a new idea in \RE. It has been studied in research on non-functional requirements. We argue that, once there are non-functional requirements in an \RP, then this \RP{} is not a subclass of the \ZJRP.}
        \item{The fourth step of the argument, in Section \ref{s:optimality-criteria-importance}, shows that optimality and non-functional requirements are central to \ASfull s engineering, and therefore, that \RPAS{} is not a subclass of the \ZJRP.}
    \end{enumerate}
    }
    \item{Part two proposes a general definition of the \RPAS. This is done in four steps:
    \begin{enumerate}
        \item{Section \ref{s:spaces} introduces new concepts, of problem and solution spaces, of criteria and parameters, and so on, for defining \RP s in general. The new concepts are motivated by the discussion in part one of the paper.}
        \item{Section \ref{s:optimal-specifications} connects the discussion of optimality to the new concepts introduced in Section \ref{s:spaces}.}
        \item{Section \ref{s:rop} introduces a new class of \RP s, called \ROPfull s, used to define the \RPAS.}
        \item{Section \ref{s:rpas} defines the \RPAS.}
    \end{enumerate}
    }
    \item{Part three, in Sections \ref{s:mathematical-optimisation}--\ref{s:expected-utility-theory}, relates the \RPAS{} to mathematical optimisation in general, to decision analysis in management science, and to expected utility theory in economics.
    }
\end{enumerate}

\section{The \ZJRPfull{} is a Minimal \RP}\label{s:minimality}
\zi{\ZJRP{} is a minimal \RP}, in the sense that if any of its parts is removed, the rest is not an interesting problem for \RE, or no problem at all. 

To see this, consider the following rewriting of the \ZJRP. The only difference from Definition \ref{d:zjrp} is that there are now labels on parts of the problem statement. Labels are used as follows: to label the statement ``it is raining'' with the label \xb{X}, we write \CLabel{it is raining}{X}.

\begin{definition}\label{d:zjrp:labeled}
\xb{\ZJRPfull{} (\ZJRP):} \xb{Given} \CLabel{a set $R$ of requirements}{R}, and \CLabel{a set $K$ of domain knowledge}{K}, \xb{find} \CLabel{a specification $S$}{S}, such that \CLabel{$S$ satisfies the following conditions}{DR}:
\begin{enumerate}
    \item{\CLabel{There is a proof of $R$ from $K$ and $S$, written $K, S \vdash R$}{\SatisfactionCondition},}
    \item{\CLabel{$K$ and $S$ are consistent, written $K, S \not\vdash \bot$}{\ConsistencyCondition}.}
\end{enumerate}
\end{definition}

There are labels on six parts of the problem statement. \xb{R} refers to the set of requirements, \xb{K} to the set of domain knowledge, and \xb{S} to the specification. \xb{DR} refers to the decision rule, that is, a rule stating what the thing to find, namely \xb{S}, needs to satisfy, in order for it to be a solution to the problem. The \SatisfactionCondition{} refers to the condition that there should be a proof of $R$ from $K$ and $S$, and the \ConsistencyCondition{} to the consistency of $K$ and $S$.

It should be fairly straightforward to notice that removing any one of the labelled parts leaves no problem at all, or no problem of interest to \RE:
\begin{itemize}
    \item{If \xb{R} is removed, then \SatisfactionCondition{} has to go too, and this remains:
        \begin{quote}
            \xb{Given} \CLabel{a set $K$ of domain knowledge}{K}, \xb{find} \CLabel{a specification $S$}{S}, such that \CLabel{$S$ satisfies the following condition}{DR}: \CLabel{$K$ and $S$ are consistent, written $K, S \not\vdash \bot$.}{\ConsistencyCondition}
        \end{quote}
    Any $S$ which is consistent with $K$ is a solution to this problem, making this an uninteresting problem for \RE{}, given that the any solution to this problem is designed independently from requirements.}
    \item{If \xb{K} is removed, then this remains:
        \begin{quote}
            \xb{Given} \CLabel{a set $R$ of requirements}{R}, \xb{find} \CLabel{a specification $S$}{S}, such that \CLabel{$S$ satisfies the following conditions}{DR}:
\begin{enumerate}
    \item{\CLabel{There is a proof of $R$ from $S$, written $S \vdash R$}{\SatisfactionCondition}}
    \item{\CLabel{$R$ and $S$ are consistent, written $R, S \not\vdash \bot$.}{\ConsistencyCondition}}
\end{enumerate}
        \end{quote}
    Note that the \ConsistencyCondition{} is changed above; another option is to remove the \ConsistencyCondition{} altogether, rather than rewrite it so that $R$ and $S$ have to be consistent. In both cases, what remains is not a relevant problem, since it says that any specification, including those defined independently from the environment conditions, will be a solution, as long as it satisfies the two conditions above.}
    \item{If the \ConsistencyCondition{} is removed, then every inconsistent specification becomes a solution. This happens if $\vdash$ is understood as the syntactic consequence relation of classical logic. This relation, then, satisfies satisfies the \zi{ex falso quod libet} proof pattern, which is that anything follows from an inconsistent set of formulas. Here, it means that if $S \vdash \bot$, then $K, S \vdash R$, whatever the content of $R$ and $K$.}
    \item{If the \SatisfactionCondition{} is removed, the result is the same as removing \xb{R}.}
\end{itemize}

\section{The \ZJRPfull{} is not the Unique \RP}\label{s:uniqueness}
The conclusion of this section will be that \zi{the \ZJRP{} is not the unique \RP}, and therefore, that \ZJRP{} is not the superclass of all \RP s. 

Section \ref{s:uniqueness:why} explains what it means for an \RP{} to be unique, and gives the main reason why it matters to know whether the \ZJRP{} is unique. Section \ref{s:uniqueness:properties} lists properties that an \RP{} can inherit from the \ZJRP. Section \ref{s:uniqueness:cases} gives three \RP s different from the \ZJRP, and discusses what properties they inherit from the \ZJRP, and in particular if they inherit all its properties. Section \ref{s:uniqueness:optimality} defines the optimality property, which is implicit in the \ZJRP. Section \ref{s:uniqueness:optimality-uniqueness} argues that there are \RP s which have a different optimality property than the \ZJRP, and therefore, are not subclasses of the \ZJRP.

\subsection{Uniqueness Matters because of Inheritance}\label{s:uniqueness:why}
Uniqueness matters, because \zi{if \ZJRP{} is the unique \RP, then all \RP s have at least the same properties as \ZJRP}. And if this is the case, then if we know how to solve \ZJRP, this should help design ways to solve \RP s in any other \RP{} class. 

Now, it may seem obvious that \RE{} involves so many different problems, such as, for example, those related elicitation and negotiation, which look nothing like the \ZJRP. And so, the conclusion from that already is that there is no unique \RP.

However, any elicitation problem, negotiation problem, and so on, which tends to arise \zi{when doing \RE}, is really a problem that arises \zi{only because a system design needs to be made or changed}. If elicitation is done without the aim of making or changing a system design, documented as a specification, then that problem is not an \RE{} problem at all.

The unique \RP, if there is one, would be the unique root of the taxonomy of \RP s. This would be \zi{the taxonomy of problems of designing systems}. Designing the system is the central problem for \RE, one that gives the motivation for solving many other problems that arise when doing \RE. These other problems, however difficult they may be, are the side effects that we have to deal with, because we are interested in designing systems.

\subsection{What Can an \RP{} Inherit from the \ZJRP?}\label{s:uniqueness:properties}
To see if an \RP{} is a subclass of another, it is necessary to define what one can inherit from the other. If an \RP{} X is a subclass of an \RP{} Y, then X will inherit all the properties of Y.

\begin{definition}\label{d:zjrp:properties}
\xb{\ZJRPfull{} properties:} The properties that an \RP{} can inherit from \ZJRP{} are motivated by the parts identified in Definition \ref{d:zjrp:labeled}. These properties are as follows:
\begin{enumerate}
    \item{\PropR: \RP{} recognises that there is a category of information which describe conditions that are desired.}
    \item{\PropK: \RP{} recognises that there is a category of information which describe conditions that hold independently from the system-to-be, and that the system-to-be has to live with.}
    \item{\PropS: \RP{} recognises that there is a category of information which describe the system-to-be.}
    \item{\PropKSR: \RP{} recognises that there are no categories of information other than those referred to by \PrR, \PrK, and \PrS{} properties, which are relevant when solving \ZJRP. In other words, all other kinds of information that may be useful are specialisations of those identified by the said properties.}
    \item{\PropSatisfaction: \RP{} requires any solution to it to be such that there is proof that, if conditions described in \PrK{} hold, and the system is implemented according to \PrS, then conditions in \PrR{} will be satisfied.}
    \item{\PropConsistency: \RP{} requires that any solution to it be such that the conditions described in the \PrK{} and \PrS{} properties are not logically inconsistent.}
    \item{\PropDR: A description of a system-to-be is a solution to the \RP{} if it satisfies the \PropConsistency{} and the \PropSatisfaction.}
\end{enumerate}
\end{definition}

\subsection{A Case of Complicated Inheritance}\label{s:uniqueness:cases}
To illustrate inheritance between \RP s, this section discusses three \RP s. 

\subsubsection{RP1} The following is the first \RP, named RP1.

\begin{quote}
    \xb{RP1:} \xb{Given} \CLabel{a set $G$ of goals}{G}, and \CLabel{a set $K$ of domain knowledge}{K}, \xb{find} \CLabel{a specification $S$}{S}, such that \CLabel{$S$ satisfies the following conditions}{DR}:
\begin{enumerate}
    \item{\CLabel{There is a proof of goals in $G$ from $K$ and $S$, written $K, S \vdash G$}{\SatisfactionCondition},}
    \item{\CLabel{$K$ and $S$ are consistent, written $K, S \not\vdash \bot$}{\ConsistencyCondition}.}
\end{enumerate}
\end{quote}

RP1 differs from \ZJRP{} in that it has no mention of requirements $R$, but says that goals in $G$ should be satisfied. However, the set of goals $G$ has the exact same role in RP1 as requirements $R$ has in \ZJRP: both are used to capture information about desired conditions that the system-to-be should satisfy. RP1 is therefore a subclass of \ZJRP, because it inherits all properties, including the \PrR{} and \PrKSR{} properties. 

\subsubsection{RP2} The second \RP{} is as follows.

\begin{quote}
    \xb{RP2:} \xb{Given} \CLabel{a set $R$ of requirements, partitioned onto mandatory requirements $R^{M}$ and non-mandatory requirements $R^{NM}$}{R}, and \CLabel{a set $K$ of domain knowledge}{K}, \xb{find} \CLabel{a specification $S$}{S}, such that \CLabel{$S$ satisfies the following conditions}{DR}:
\begin{enumerate}
    \item{\CLabel{There is a proof of mandatory requirements in $R^{M} \subseteq R$ from $K$ and $S$, written $K, S \vdash R^{M}$}{\SatisfactionCondition},}
    \item{\CLabel{$K$ and $S$ are consistent, written $K, S \not\vdash \bot$}{\ConsistencyCondition}.}
\end{enumerate}
\end{quote}

The difference is between RP2 and \ZJRP{} is that the set of requirements $R$ is partitioned onto mandatory and non-mandatory requirements in RP2. A solution therefore does not need to satisfy all requirements in $R$, but a subset thereof. RP2 is equivalent to \ZJRP{} when all requirements in $R$ are mandatory. 

There are two ways of looking at inheritance between RP2 and \ZJRP. One is to say that RP2 specialises the concept of requirement onto mandatory and non-mandatory requirement, and rewrites the \SatisfactionCondition{} accordingly. The other is that RP2 is obtained by taking that $R$ in \ZJRP{} includes only mandatory requirements, and saying that there are other, non-mandatory requirements which remain outside \ZJRP; in this second case, RP2 is the same problem as \ZJRP, with the added set of non-mandatory requirements, which remain unrelated to the specification, and thereby not a factor that influences the design of the specification. 

In both cases, RP2 looks like a subclass of \ZJRP, because non-mandatory requirements play no role in the problem or the solution. RP2 thereby inherits all properties of \ZJRP, and adds two properties: one is that any requirement is either mandatory or non-mandatory, and the other that a solution should satisfy all mandatory requirements.

Returning to the more general discussion, recall that in Section \ref{s:minimality}, it was argued that the \ZJRP{} is minimal by looking at what remains after some part of it is removed. 

There is a clear correspondence between parts of the \ZJRP{} in Definition \ref{d:zjrp:labeled} and the properties in Definition \ref{d:zjrp:properties}. 

Due to that correspondence, it follows that \zi{if a part is removed, then what remains will fail to satisfy all the \ZJRP{} properties}. Therefore, the set of properties in Definition \ref{d:zjrp:properties} is minimal.

\subsubsection{RP3} It was straightforward to determine what RP1 and RP2 inherited from \ZJRP. RP3 is a more complicated case.

\begin{quote}
    \xb{RP3:} \xb{Given} \CLabel{a set $R$ of requirements, partitioned onto mandatory requirements $R^{M}$ and non-mandatory requirements $R^{NM}$}{R} and \CLabel{a set $K$ of domain knowledge}{K} , \xb{find} \CLabel{a specification $S$}{S}, such that \CLabel{$S$ satisfies the following conditions}{DR}:
\begin{enumerate}
    \item{\CLabel{There is a proof of mandatory requirements in $R^{M} \subseteq R$ from $K$ and $S$, written $K, S \vdash R^{M}$}{\SatisfactionCondition},}
    \item{\CLabel{$K$ and $S$ are consistent, written $K, S \not\vdash \bot$}{\ConsistencyCondition},}
    \item{\CLabel{There is no other specification $S^{\prime}$ which satisfies both the \SatisfactionCondition{} and the \ConsistencyCondition, and in addition satisfies more of the non-mandatory requirements in $R^{NM}$ than does $S$}{\OptimalityCondition.}}
\end{enumerate}
\end{quote}

RP3 partitions requirements onto mandatory and non-mandatory. In RP2, the non-mandatory requirements had no influence on which specification will be the solution. In RP3, the non-mandatory requirements appear in the \OptimalityCondition, that a specification has to satisfy in order to be the solution.

This suggests that the solution concept in RP3 is a subclass of the solution concept in RP2. In RP2, a solution is any specification which satisfies the \SatisfactionCondition{} and the \ConsistencyCondition, while in RP3, the specification also has to satisfy the \OptimalityCondition. In other words, the extension of the RP3 solution concept is a subset of the RP2 solution concept.

\subsection{Optimality in the \ZJRPfull}\label{s:uniqueness:optimality}
The \OptimalityCondition{} in RP3 is significantly different from the \SatisfactionCondition{} and the \ConsistencyCondition, because \zi{the \SatisfactionCondition{} and \ConsistencyCondition{} are verified on a single specification, and it does not matter what other specifications there may be, while to verify if the \OptimalityCondition{} is satisfied, it is necessary to compare two or more specifications}.

It is therefore not possible to check if the \OptimalityCondition{} in RP3 is satisfied, when looking at one specification independently from others.

If we found a single specification $S$, which satisfies the \SatisfactionCondition{} and the \ConsistencyCondition, then in absence of at least one other specification $S^{\prime}$ with which to compare $S$ in terms of how many non-mandatory requirements they satisfy, there will be no justification to the claim that $S$ satisfies the \OptimalityCondition.

The \ZJRP{} has its own notion of optimality, which is implicit in the \PropDR. 

To see it, suppose that there are three specifications $S_{1}$, $S_{2}$, and $S_{3}$, and that they all satisfy the \SatisfactionCondition{} and the \ConsistencyCondition{} for the same set of requirements $R$ and the same domain knowledge $K$. Which of the three specifications is the optimal one?

The \PropDR{} says that a specification is the solution if it satisfies the \SatisfactionCondition{} and the \ConsistencyCondition. As there is no other property that a specification needs to satisfy to be a solution, the only remaining conclusion is that \zi{any specification that satisfies the \SatisfactionCondition{} and the \ConsistencyCondition{} is optimal}.

To make explicit the notion of optimality in \ZJRP, we add the following property to the \ZJRP.

\begin{definition}\label{d:zjrp:propopt}
\xb{\PropOpt{} for the \ZJRP:} \RP{} recognizes that if there are more than one description of the system-to-be, all of which satisfy the \PrSatisfaction{} and \PrConsistency{} Properties, then they are all equally desirable.
\end{definition}

This leads to the following revision of the properties that an \RP{} can inherit from the \ZJRP.

\begin{definition}\label{d:zjrp:properties-revised}
\xb{\ZJRPfull{} properties (revised):} The properties that an \RP{} can inherit from the \ZJRP{} are \PrR, \PrK, \PrS, \PrKSR, \PrSatisfaction, and \PrConsistency{} Properties from Definition \ref{d:zjrp:properties}, the \PropOpt{} from Definition \ref{d:zjrp:propopt}, and the following property:
\begin{itemize}
    \item{\PropDR: A description of a system-to-be is a solution to the \RP{} if it satisfies the \PrConsistency, \PrSatisfaction, and \PrOpt{} Properties.}
\end{itemize}
\end{definition}


%
\subsection{How are Optimality and Uniqueness related?}\label{s:uniqueness:optimality-uniqueness}
Optimality is important, because it is related to uniqueness in the following way: 

\zi{In order to establish if a specification is the optimal specification and therefore the solution to the \RP, it is necessary to compare specifications; any comparison of specifications requires having in the \RP{} the information about comparison criteria; such information is not part of requirements, domain knowledge, or of the specification in the \ZJRP, which violates the \PropKSR, and therefore, the \ZJRP{} is not unique.}

To understand the above, suppose that in a concrete systems engineering project, we have the set $R$ of requirements and $K$ of domain knowledge. In RP1, we simply called every requirement a goal, so that \zi{both the RP1 and the \ZJRP{} have the same set of specifications, each of which can be a solution}. 

It makes sense therefore to conclude that the RP1 is a subclass of \ZJRP, or that they are equivalent \RP s, because for any given $R$ and $K$, \zi{solving the RP1 or the \ZJRP{} will involve choosing one solution from the same set of potential solutions}. Moreover, in absence of comparison criteria in either the \ZJRP{} and RP1, we can choose any of the specifications as the solution.

For simplicity, the set of all specifications that we can choose one solution from, when solving an \RP{}, will be called the \xb{\SolutionSpace} of that \RP.

So the \SolutionSpace{} is the same for RP1 and \ZJRP, for the same given sets of requirements and domain knowledge.

Despite the similarities between RP2 and \ZJRP, their \SolutionSpace s are different. In \ZJRP, all requirements in a given set $R$ of requirements must be satisfied. In RP2, that same set is partitioned onto mandatory $R^{M}$ and non-mandatory $R^{NM}$ requirements. It follows that the \SolutionSpace{} of RP2 for given requirements $R$ and domain knowledge $K$ is the same as the \SolutionSpace{} of \ZJRP{} defined over those same requirements and domain knowledge \zi{only if $R^{M} = R$ and $R^{NM} = \emptyset$}. Instead, \zi{if any member of $R$ is in $R^{NM}$, then it is by definition of RP2 not in $R^{M}$, and consequently, the \SolutionSpace{} of RP2 will not include the same specifications that the \SolutionSpace{} of the corresponding \ZJRP{} would.}

If an \RP{} X is a subclass of another \RP{} Y, then for the same given requirements and domain knowledge, every solution to X should also be a solution to Y. This is not the case in RP2.

The RP3 is not a subclass of the \ZJRP{} for the same reason: as soon as some members of $R$ are in $R^{NM}$, the \SolutionSpace s of RP3 and \ZJRP{} will differ, as some specifications that can be solutions to the RP3 will fail to satisfy the non-mandatory requirements in $R^{NM}$ and therefore cannot be in the \SolutionSpace{} of the \ZJRP{} defined over the same set of requirements $R$ and the same domain knowledge $K$.

Now, it is fair to observe that RP2 is an odd \RP, because there seems to be no role for non-mandatory requirements in it. But this same observation cannot be made for RP3, where non-mandatory requirements serve to define the criterion for the comparison of specifications in the \SolutionSpace; that criterion is given in the \OptimalityCondition{} of RP3. 

It is interesting to note that the conclusion we got here is counter-intuitive. To make RP3, we did three operations: (i) we specialised requirements onto mandatory and non-mandatory ones, (ii) revised the \SatisfactionCondition, so that it reflects the idea that only mandatory requirements must be satisfied, and (iii) added the \OptimalityCondition. The three operations are non-controversial; we could not do (i) without also doing (ii), and doing (i) also made no sense without doing (iii). 

The result is counter-intuitive, because the operation (i) looks simply like the specialisation of a concept that is already there in the \ZJRP, and (iii) as just us asking that every solution satisfies an additional property, and so a specialisation of the \ZJRP{} solution concept. If we look at each of these operations in isolation, they look like we are merely adding detail to what the \ZJRP{} already had.

But together, the three operations resulted in a different problem to solve, because by solving RP3, we can get solutions to RP3 which are not solutions to \ZJRP. Hence, RP3 cannot be a specialisation of the \ZJRP.

\section{Non-functional Requirements as Comparison Criteria}\label{s:non-functional-requirements}
Non-functional requirements are an important source of comparison criteria. This section will argue that, \zi{if} there are non-functional requirements in an \RP, \zi{and} we extract from them the criteria for the comparison of specifications, \zi{and} we want to find the optimal specification according to those criteria, \zi{then} the \RP{} is not a subclass of the \ZJRP.

Suppose that stakeholders give us non-functional requirements, also called quality requirements \cite{boehm1976quantitative, mylopoulos1992representing}. For an ambulance system, an example can be the requirement that ``ambulances should quickly arrive at incident locations''; denote this requirement \req{r1}. There is no universal criterion or industrial standard for just how much time amounts to ``fast'' in this requirement. 

If we wanted to keep solving the \ZJRP{} in the presence of non-functional requirements, we could do this by replacing each non-functional requirement by a variant, the satisfaction of which is binary. For example, ``on average, ambulances should arrive to incident locations within 10 minutes''; denote this \req{r2}. And we could then have such $K$ and $S$, that we can prove \req{r1} from them. All looks as if we are still solving the \ZJRP. 

But this transformation of \req{r1} misses the point. While \req{r1} looks like a requirement, its role in the \RP{} is completely different than that of \req{r2}. 

The non-functional requirement \req{r1} \zi{serves as a criterion for the comparison of alternative specifications, because it states a preference relation}: by saying that ambulances should quickly arrive, it states that, when given two systems, one in which ambulances arrive slower, that one will -- over this criterion only -- be strictly less desirable than the system in which ambulances arrive faster. This is not at all the same as saying that ambulances should arrive within 10 min, since in that case, \zi{any} system which satisfies this is good enough, while in the former case, only that system which achieves -- among all those cosnidered -- the shortest time for the ambulance to arrive, is good enough (again, if this criterion alone is considered, because when there are many criteria, perhaps some other criterion will have more importance).

Different specifications, each associated to a different design of the system-to-be, may result in different average time for an ambulance to arrive at the incident location. Some specifications will result in systems that will be faster, others slower. 

There is, therefore no sense in placing \req{r2} in $R$, because how fast one specification is (or, to be precise, how fast we expect the resulting system to be), is relative between specifications. We cannot prove it from $K$, and $S$ for \zi{one} specification, because we have to take into account other, alternative, specifications. 

Suppose that we do not accept the arguments above, and we want to do something to non-functional requirements in order to still keep solving the \ZJRP. Here are some possible attempts to repair the situation:
\begin{enumerate}
    \item{We could rewrite \req{r1} as \req{r3}, with \req{r3} denoting the statement ``specification $S$ is the specification with the lowest time for an ambulance, on average, to arrive at an incident location'', and put \req{r3} in $R$. But notice that \req{r3} is a rather odd requirement, since it \zi{is about the relationship between different specifications}. This is also a practical problem, as it is not clear how we could prove \req{r3} from $K$ and $S$, unless $K$ and/or $S$ also are about, or somehow mention other specifications.}
    \item{We could say that \req{r1} is not a member of $R$, but stays outside $K$, $S$, and $R$, and that, to enable the comparison of specifications, we should add a variable, denote it $q$, for ``average time to arrive at incident location'', and that we should simulate, or estimate otherwise the value of $q$ for each specification. If the various non-functional requirements result in a set $Q$ of such variables, we could revise \ZJRP{} by requesting that, for each specification $S_{i}$, this holds: 
    \begin{equation}
        K, S_{i} \vdash R, Q_{i} 
    \end{equation}
    where $Q_{i}$ is the set of value assignments to all variables in $Q$, produced by the specification $S_{i}$. We could then compare different specifications by the values they assign to variables, whereby we chose the variables to quantify level of satisfaction of non-functional requirements; for example, we could assume that there is a scale for time for \req{r1}, where values indicate the average time to arrive at an incident location.}
    \item{We can replace syntactic consequence $\vdash$ with another relation; if we denote the new relation with $\nmdash$, we would need to define it in such a way that we have:
    \begin{equation}
        K, S_{i} \nmdash R
    \end{equation}
    if and only if (i) $R$ is provable from $K$ and $S_{i}$, \zi{and} (ii) there is no other specification $S_{j}$ that better satisfies the non-functional requirements. Defining ``better'' requires us to define a way to aggregate, for each specification, the levels at which it satisfies all non-functional requirements, and then compare that aggregate score with those of all other specifications. The best specification would be the one with the highest score.}
\end{enumerate}

In each of the three approaches above, we made significant changes to \ZJRP, in order to accommodate non-functional requirements. That is, we made new problems, different from the \ZJRP:
\begin{itemize}
    \item{In the first approach, we revised the non-functional requirement \req{r1} as \req{r3} and placed \req{r3} in $R$. However, \req{r3} is about other specifications, yet \ZJRP{} is about conditions that a single specification should satisfy, independently from others.}
    \item{In the second approach, we had to add $Q_{i}$, the assignment of values to measures of non-functional requirements. It is not a problem to have $Q_{i}$ as a subset of $R$. However, if we want to choose the specification $S_{i}$ which gives the most desirable $Q_{i}$, then we are not solving the \ZJRP, because the part \xb{DR} in \ZJRP{} does not say that we should choose the most desirable specification which satisfies the \SatisfactionCondition{} and the \ConsistencyCondition. It actually says nothing about how desirable the specification ought to be, relative to other specifications which also satisfy these conditions.} 
    \item{Finally, in the third approach, we replaced the provability condition with a relation $\nmdash$, which has to take into account the level to which non-functional requirements are satisfied in all considered specifications.}
\end{itemize}

To summarise, while \ZJRP{} is minimal, it is not unique. As soon as there is information about criteria for the comparison of specifications in the \SolutionSpace{} \zi{and we are interested in choosing the specification which best satisfies these non-functional requirements (whatever ``best'' may mean precisely in a specific project)}, then we are solving a different problem than \ZJRP.

\section{Optimality and Comparison are Central to \ASfull s}\label{s:optimality-criteria-importance}
At this point, the important conclusion is that the \ZJRP{} is not the superclass of \RP s, in which we are interested in finding the optimal specification, and we have information that lets us compare specifications.

The claim in this section is that \zi{\RPAS{} is not a subclass of the \ZJRP, because optimality and comparison play a central role in it}: namely, adaptation amounts to the switching between alternative ways of satisfying requirements, and therefore, each time the system needs to adapt, it needs to compare alternative ways of adapting, and choose the optimal one, among those that are available.

To further clarify this discussion, the rest of this section uses a trivial and hypothetical example, but sufficient to support these claims.

\subsection{Adaptation as Switching}\label{s:optimality-criteria-importance:switching}
In this example, by \zi{qualitative requirement}, we mean a requirement for which we say that it is either satisfied or not, not satisfied to some extent. By \zi{quantitative variable requirement}, we mean a requirement that assigns a desirable value or range of values to a variable, which is not binary. 

We have only two qualitative requirements \req{rA} and \req{rB} to satisfy. We know that we can satisfy \req{rA} by implementing one of five different functionalities, denoted \req{rAF1} to \req{rAF5}, and \req{rB} by implementing another five different functionalities, denoted \req{rBF1} to \req{rBF5}. For simplicity, let all ten functionalities be different and not related in terms of refinement or parthood, that is, they are neither more detailed variants of one another, nor parts of one another. 

With two qualitative requirements and 5 functionalities satisfying each, there are 25 combinations of the 10 functionalities. But, some functionalities are not compatible. This means that we cannot make a system which includes them both. Some combinations of functionalities therefore do not give a specification which satisfies both \req{rA} and \req{rB}. 

\begin{figure}[t]
	\centering
	\includegraphics[height=70mm]{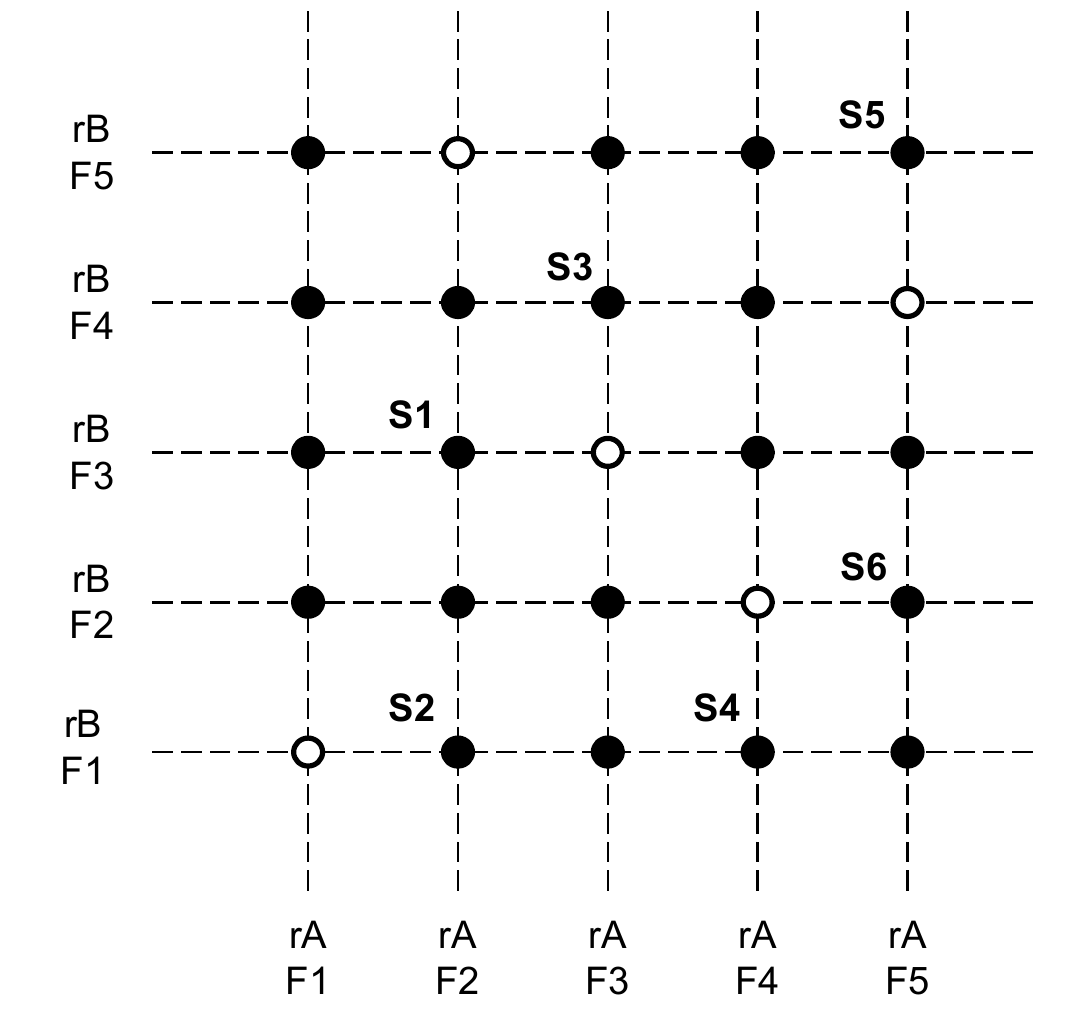}
\caption{\req{rA} and \req{rB} are two qualitative requirements that both need to be satisfied by a system. \req{rAF1} to \req{rAF5} are alternative functionalities that satisfy \req{rA}, while \req{rBF1} to \req{rBF5} are alternative functionalities that satisfy \req{rB}. Filled circles are combinations of functionalities that satisfy both \req{rA} and \req{rB}, and are thereby alternative specifications of a system; empty circles denote incompatible combinations of functionalities, and are not specifications.}
\label{f:ex:specifications}
\end{figure}

The part of the example introduced so far can be drawn as in Figure \ref{f:ex:specifications}. There, filled circles are specifications, that is, combinations of functionalities that satisfy both \req{rA} and \req{rB}. Empty circles are incompatible combinations of functionalities, and since we cannot make a system that has those functionalities, these empty circles are not specifications.

If our problem was to find one combination where functionalities are compatible, and which satisfies both \req{rA} and \req{rB}, then this can be any one of the 20 specifications in Figure \ref{f:ex:specifications}.

\begin{figure}[t]
	\centering
	\includegraphics[height=70mm]{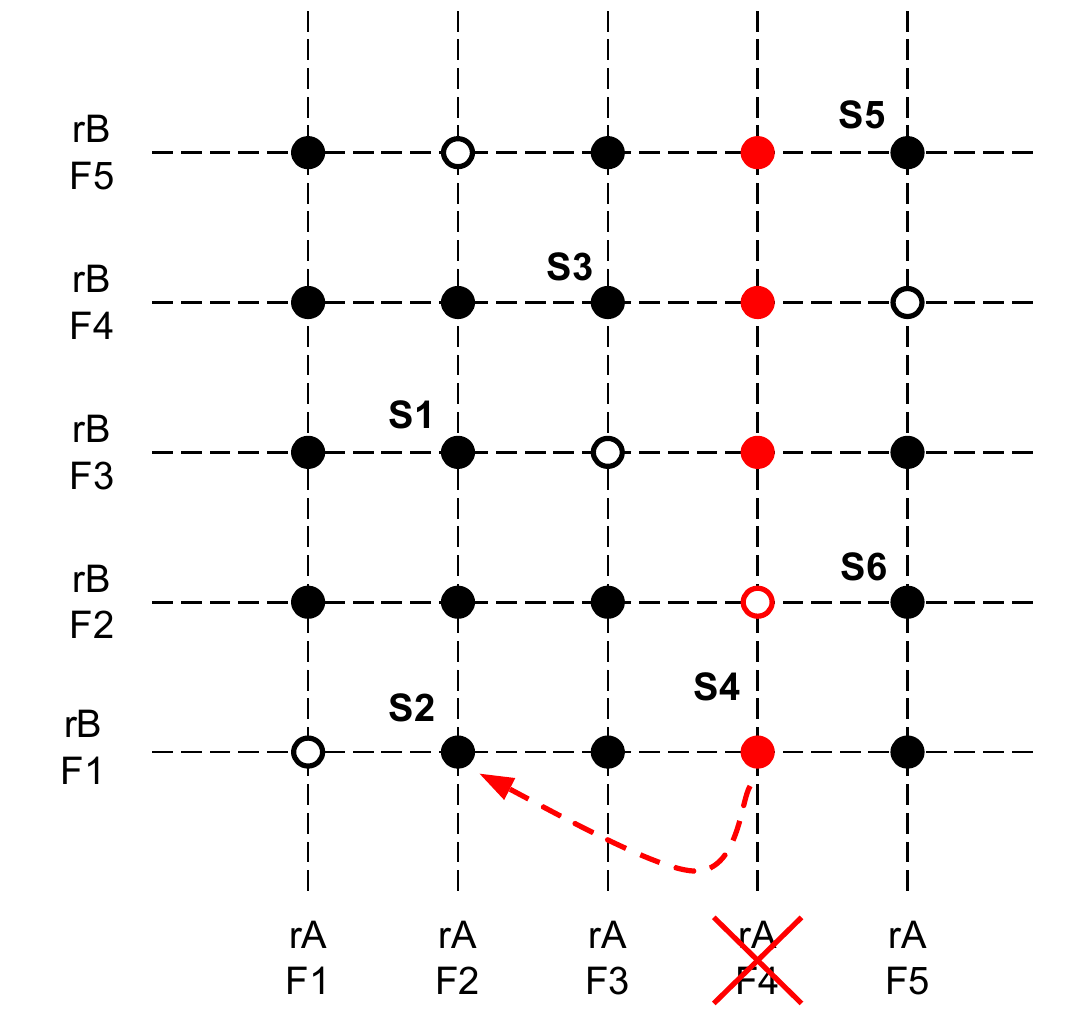}
\caption{System runs according to specification S4. Then, functionality \req{rAF4} fails, and the system activates functionality \req{rAF2} instead, switching thereby from specification S4 to specification S2.}
\label{f:ex:specifications-switch}
\end{figure}

Consider adaptation to the failure of a functionality. If we were to design a system according to one specification among those in Figure \ref{f:ex:specifications}, suppose that we chose, for example, S4, so that the system has functionalities \req{rAF4} and \req{rBF1}. If \req{rAF4} fails, the system would stop working as expected. If it were an \ASfull, then it would be designed so as to switch to another functionality at run-time, for example, from \req{rAF4} to \req{rAF2}. From the perspective of design-time, this amounts to a switch from the design given in specification S4, to that in S2. This is illustrated in Figure \ref{f:ex:specifications-switch}.

Now, assume that we have two quantitative variable requirements to satisfy, in addition to \req{rA} and \req{rB}. The first relates to scalability and the second to how the system compares to its competitors. Let \req{Var1} be the variable in the first, and Var2 in the second quantitative variable requirement. \req{Var1} can be ``number of users that can simultaneously use the system'' and Var2 ``number of products available for purchase''. 

\begin{figure}[t]
	\centering
	\includegraphics[height=70mm]{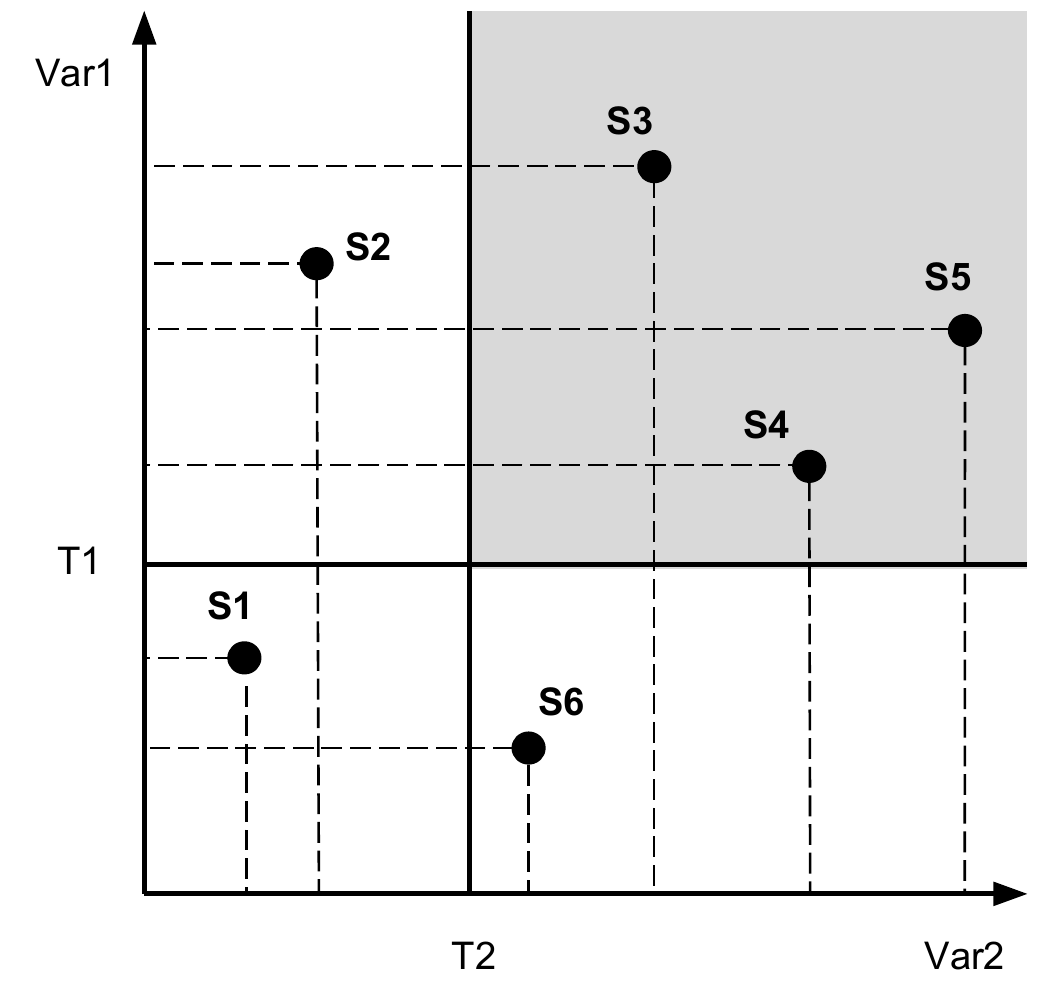}
\caption{Variables \req{Var1} and \req{Var2} quantify the level of satisfaction of two quantitative variable requirements. Hypothetical simulations of specifications S1 to S6 yield values show in the figure. T1 is the threshold value for \req{Var1}, and T2 for \req{Var2}.}
\label{f:ex:mapping}
\end{figure}

We prefer large to small values for both \req{Var1} and \req{Var2}, and we cannot accept values that are below some threshold. This is drawn in Figure \ref{f:ex:mapping}, where T1 is the threshold for \req{Var1} and T2 for \req{Var2}, so that the shaded area shows all acceptable combinations of \req{Var1} and \req{Var2} values.

If the system were running according to S4, then it satisfies all four requirements, as its values over \req{Var1} and \req{Var2} are above their respective thresholds. If \req{rAF4} fails, the system would need to switch from S4 to either S3 or S5 in order to still satisfy all four requirements; if it switched to S2, it would satisfy \req{rA}, \req{rB}, and the requirement on \req{Var1}, but not the requirement on \req{Var2}.

As long as the system can switch from one specification to \zi{any} other specification, provided that the latter satisfies all requirements and domain knowledge, then we can capture with the \ZJRP{} the problem of designing that system's specification.

\subsection{Switching and Optimality}\label{s:optimality-criteria-importance:switching-and-optimality}
Switching to \zi{any} specification fails to capture that \zi{not all alternative specifications, that the system can switch to, are equally desirable}. The goal is to switch to the one that is optimal with regards to the requirements and domain knowledge, that the system is sensing.

This concern with whether the specification to switch to, is the optimal specification among those that we can switch to, clearly distinguishes \RPAS{} from \ZJRP.

We said that we prefer higher values of \req{Var1} and \req{Var2}, or in other words, we said that we have non-functional requirements which can be interpreted as suggesting that we prefer higher values of these variables. To make this more precise, we need to say which combinations of values we prefer over others. Since the region above the thresholds T1 and T2 is large, it is interesting to indicate (i) the shape of indifference curves in that region, and (ii) the direction where these indifference curves are over more desirable combinations of \req{Var1} and \req{Var2} values. 

\begin{figure}[t]
	\centering
	\includegraphics[height=70mm]{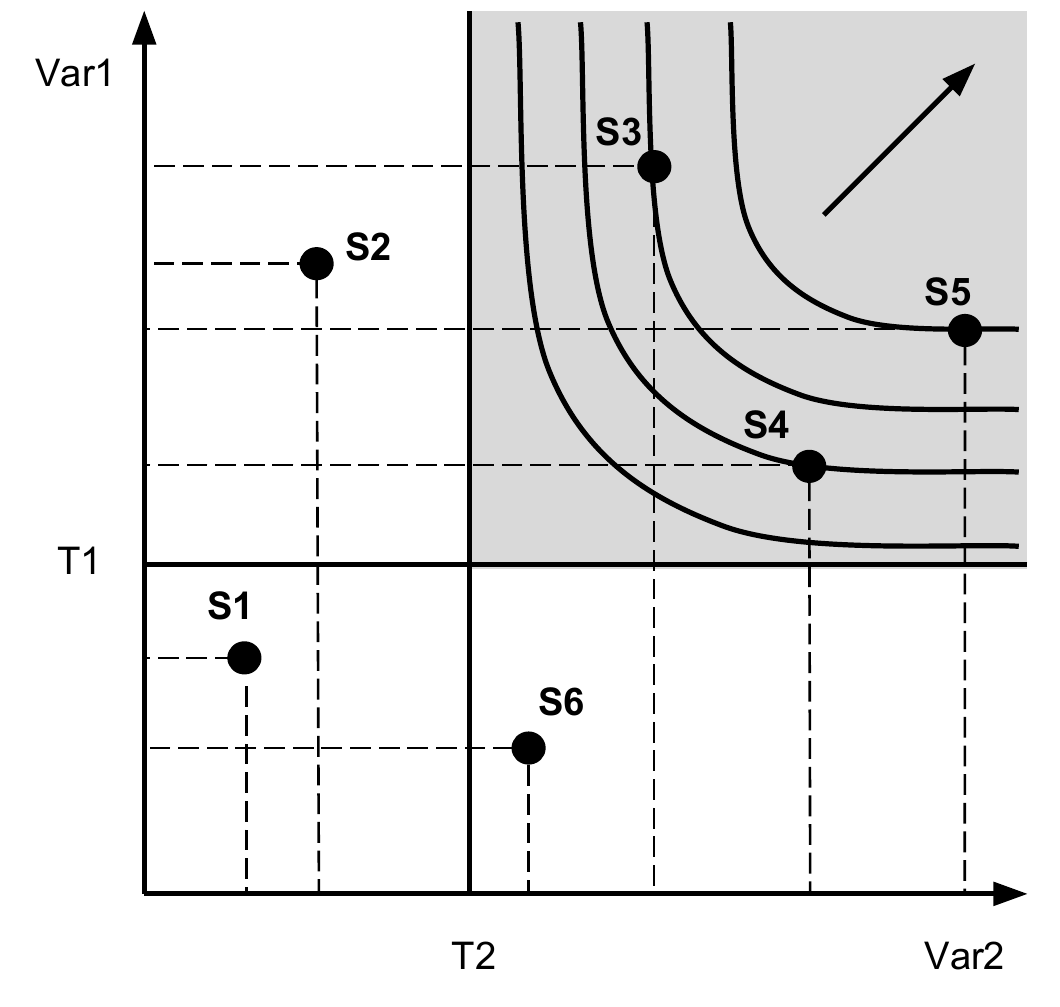}
\caption{Hypothetical indifference curves over value combinations of \req{Var1} and \req{Var2}. One indifference curve includes all \req{Var1} and \req{Var2} value combinations that are equally desirable. For example, any specification on the same indifference curve as S3 is equally desirable as S3. The arrow indicates the direction in which value combinations are more desirable. Therefore, S5 is preferred to S3, and S3 is preferred to S4.}
\label{f:ex:indifference-curves}
\end{figure}

Figure \ref{f:ex:indifference-curves} shows hypothetical indifference curves in the region above thresholds T1 and T2. Each indifference curve \zi{is the set of Var1 and \req{Var2} value combinations that are equally preferred}. So every specification that has \req{Var1} and \req{Var2} values on the same indifference curve as S3 is equally preferred to S3. The arrow indicates the direction where \req{Var1} and \req{Var2} value combinations are preferred, so that any specification on the indifference curve with S5 is strictly preferred to any specification on the indifference curve with S3.

Having clarified with indifference curves what we mean by preference for higher \req{Var1} and \req{Var2} values, we now go back to Figure \ref{f:ex:specifications}. There, we had 20 alternative specifications. If we want to see them as subsets of S in \ZJRP, they are 20 alternative \xb{configurations} of the same system. Moreover, we said that adaptation amounts to moving from one to another of these configurations, based on sensory input of the system, and feedback mechanisms that indicate what configuration to switch to. We also said that all this can be captured in \ZJRP.

By adding the two quantitative variable requirements, with their \req{Var1} and \req{Var2}, we had restricted the set of acceptable configurations to some subset of the 20 shown in Figure \ref{f:ex:specifications}. 

Given several configurations, all above T1 and T2 thresholds, and all satisfying \req{rA} and \req{rB}, which one should we choose? 

The answer is simple: given the indifference curves, we should choose any configuration which is on the the most desirable indifference curve. More generally, we want the system to switch, every time it needs to adapt, to the configuration that is the most desirable, among those that are feasible. 

This is not to say that we cannot capture also this notion of optimality in \ZJRP. For example, we can have as the only member of R in \ZJRP, the proposition that there should be no feasible configuration which is on a more desirable indifference curve than the chosen configuration. We can, therefore see \RPAS{} as a subclass of the \ZJRP, although doing so seems rather odd, for there were no considerations of configurations, adaptation, preference, uncertainty, or optimality in defining the \ZJRP.

\section{Requirements Problem and Solution Spaces}\label{s:spaces}
We start with two simple definitions; they clarify what we will mean by the terms \Specification{} and \Solution{} in the rest of the paper. 

\begin{definition}\label{d:specification}
\xb{\Specification:} A Specification is a description of a design of a system.
\end{definition}

\begin{definition}\label{d:solution}
\xb{\Solution:} In a given \RP, a \Solution{} is a \Specification{} such that, if we chose to commit to making the system according to that \Specification, then we consider that we have solved that \RP.
\end{definition}

Another way to think about the Solution and Specification, is that the Specification is a candidate Solution in an RP that we are solving.

Hereafter, and in general in the paper, if we define a term, then we will capitalise it throughout. The capitalised term should be read as it was defined.

The conclusion that not all \RP s are subclasses of the \ZJRP, and that \RPAS{} is also not a subclass of the \ZJRP, means that there are many \RP s that we may want to define and solve, and that each may have many \Specification s, some or one of which is the optimal one, and thereby the \Solution.

The goal now is to narrow down these ideas, and we will do this by defining the notions of \ProblemSpace{} and \SolutionSpace. If we agree on what these spaces are, then it will be easier to define the \RPAS.

\subsection{\ProblemSpace}\label{s:spaces:problem-space}
We call it the \zi{\ProblemSpace}, because it has some number of dimensions, and each dimension corresponds to something that we can evaluate a \Specification{} for. 

For example, if we have a \ZJRP{} instance with one requirement $|R| = 1$ and no domain knowledge $|K| = 0$, then the \ProblemSpace{} would be one-dimensional, a line, where each point refers to a level of satisfaction of the single requirement in $R$. If that requirement had a scale of satisfaction with 10 levels, then there would be exactly 10 points in the \ProblemSpace. If it had two levels of satisfaction - satisfied or failed - then the \ProblemSpace{} would have two points only. If the level of satisfaction was a real number between some maximum and minimum, then there would be an infinite number of points in the \ProblemSpace.

In an \RP{} which satisfies the \PrR{} and the \PrK{} properties, we can evaluate if a Specification satisfies a requirement, so that members of $R$ would induce dimensions in the problem space. But we can also evaluate if the \Specification{} satisfies or fails constraints from $K$, which is why members of $K$ would also induce dimensions in the problem space.

To remain general, we need to avoid being constrained by the \ZJRP. In particular, we want to be independent from the properties \PrK{} and \PrR, that is, from the categorization that the \ZJRP{} imposes on criteria that \Specification s are evaluated against. We do this by introducing the concept of \Criterion{} in the definition of the \ProblemSpace{} concept.

\begin{definition}\label{d:problem-space}
\xb{\ProblemSpace:} Set of points, where each coordinate of each point corresponds to the value of a \Criterion.
\end{definition}

\begin{definition}\label{d:criterion}
\xb{\Criterion:} A variable, such that (i) we can establish its value for each \Specification, and (ii) some of its values are more desirable than others, regardless of which \Specification{} is being evaluated.
\end{definition}

Various kinds of information used when solving an \RP{} can produce \Criteria{} for a \ProblemSpace. The obvious example are requirements that are either satisfied or not. For each of those, we have one \Criterion. non-functional requirements are more complicated, since it can be hard to find suitable variables to measure their level of satisfaction. Such variables would correspond to \Criteria. 

\Criteria{} do not come from requirements only. Domain knowledge may indicate laws that the system-to-be should comply with, and each norm from the law may give a \Criterion. Each of those \Criteria{} would have two values, does comply and does not comply.

There is a nuance to keep in mind: some requirements, for example, may be refinements of other requirements, so that there can be relations between the value of different \Criteria{} for the same \Specification. For example, the value of a \Specification{} on one \Criterion{} may be fully determined by values of that \Specification{} on other \Criteria. The following example illustrates this.

In ambulance services, suppose that we have this statement: ``Ambulances arrive at their incident locations''.

We will take this to be a requirement, and abbreviate it with \req{R(AmbArrive)}, where \req{AmbArrive} refers to the statement above, and \req{R} that this statement is a requirement.

We can make this requirement more detailed, by saying that it will be satisfied if these more specific requirements are satisfied: \req{R(IdentAmb)} for ``Identify available ambulances'', \req{R(ChooseAmb)} for ``Choose ambulance to dispatch'', \req{R(AssignAmb)} for ``Assign ambulance to incident'', \req{R(MobilizeAmb)} for ``Dispatch the ambulance to the incident location'', \req{R(ConfirmMob)} for ``Confirm that the ambulance was dispatched''.

In other words, we refined \req{R(AmbArrive)} onto five other requirements. As this means that if all five are satisfied, then \req{R(AmbArrive)} is satisfied, it also means that in the \ProblemSpace: 
\begin{itemize}
    \item{there is a Criterion for each of the six requirements above,}
    \item{each of these Criteria allows two values, satisfied or not satisfied (this is the case if we do not want to allow degrees of satisfaction for these requirements, but we see them as either satisfied or not),}
    \item{there is a function that ties the satisfaction of the five more specific requirements with the refined requirement, in that the latter is satisfied only if all five are satisfied. This means that the value of the Criterion corresponding to R(AmbArrive) is function of values for Criteria for the five other requirements.}
\end{itemize}

\begin{figure}[t]
	\centering
	\includegraphics[height=70mm]{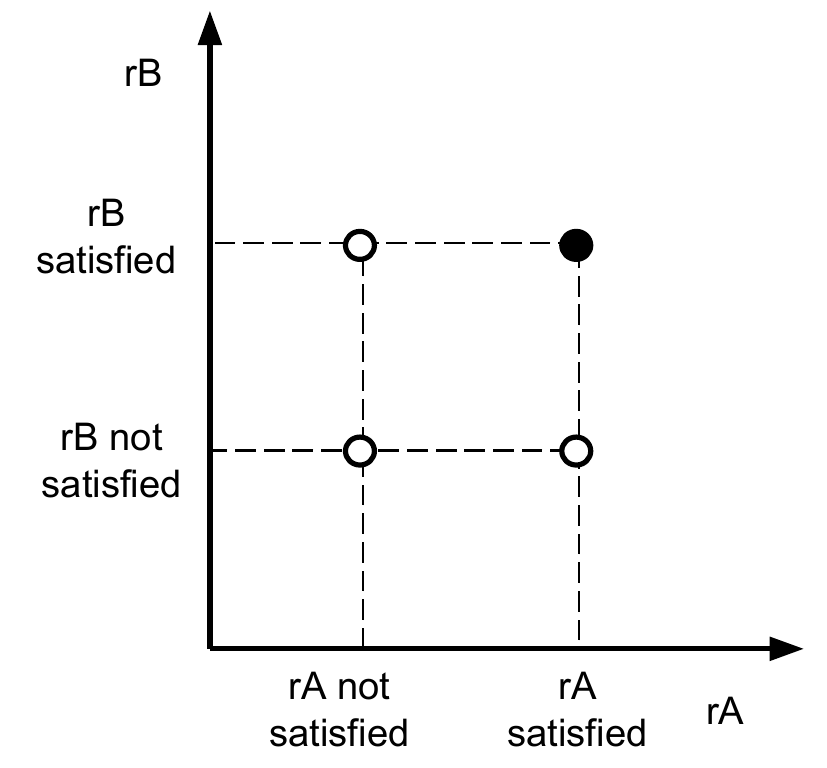}
\caption{Problem space defined by two requirements \req{rA} and \req{rB}, both with two levels of satisfaction. The satisfaction of one is independent from the satisfaction of the other. The black circle is a position in the \ProblemSpace{} where both requirements are satisfied.}
\label{f:ex:problem-space-ra-rb}
\end{figure}

In the example in Section \ref{s:optimality-criteria-importance:switching} we had two requirements, \req{rA} and \req{rB}. We evaluated each as either satisfied or not. It follows that our \ProblemSpace{} there has two \Criteria, \req{r1} and \req{r2}. If we further assume that the satisfaction of one requirement is independent from the satisfaction of the other, then any \Specification{} can correspond one of four points in the \ProblemSpace. This is illustrated in Figure \ref{f:ex:problem-space-ra-rb}.

A more complicated \ProblemSpace{} may involve non-functional requirements, which give \Criteria{} that can take a real value from some range. Each \Specification{} evaluates to one value of that \Criterion, so that the number of positions that \Specification s can take is infinite. 

\begin{figure}[t]
	\centering
	\includegraphics[height=70mm]{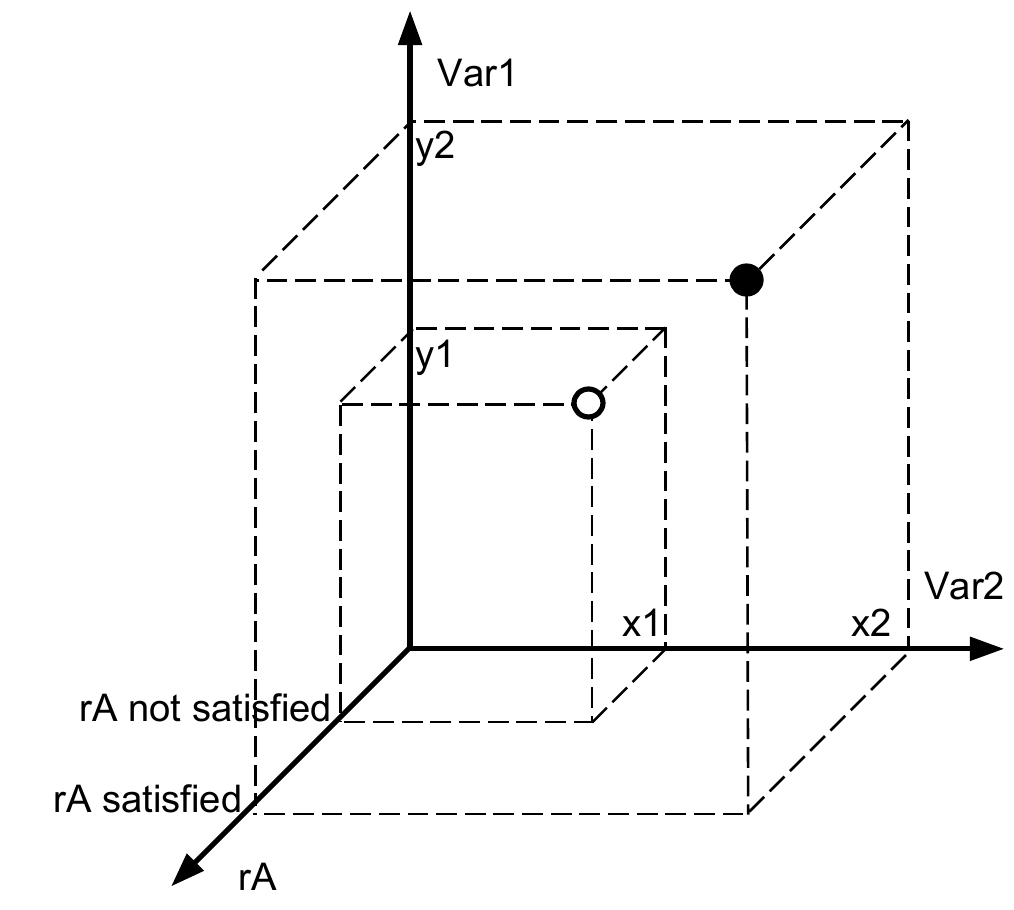}
\caption{\ProblemSpace{} defined by the binary requirement \req{rA} and two non-functional requirements from Figures \ref{f:ex:specifications} and \ref{f:ex:mapping}. Black circle is a point where \req{rA} is satisfied, while the empty circle is a point where \req{rA} fails.}
\label{f:ex:problem-space-ra-var1-var2}
\end{figure}

Figure \ref{f:ex:problem-space-ra-var1-var2} illustrates the \ProblemSpace{} defined by one binary requirement and two variables quantifying the degree of satisfaction of non-functional requirements. There, the \ProblemSpace{} amounts to two planes, one where a \Specification{} has the coordinates $(\zi{rA not satisfied}, x1, y1)$ and the other where a Specification's coordinates are $(\zi{rA satisfied}, x2, y2)$; $x1$ and $x2$ are values of \req{Var1}, $y1$ and $y2$ of \req{Var2}.

\subsection{\ProblemInstance s}\label{s:spaces:problem-instances}
It is important to observe that \Criteria{} define a \ProblemSpace, not a single \RP{} to solve. 

If we have a \ProblemSpace{} made of $n$ \Criteria, we can choose exactly one \RP{} to solve by choosing one value of each \Criterion. By choosing a point in the \ProblemSpace, we have identified the exact requirements, domain knowledge, etc., that any \Specification{} should satisfy. We capture this in our terminology by adding the following.

\begin{definition}\label{d:problem-instance}
\xb{\ProblemInstance:} A \ProblemInstance{} in a \ProblemSpace{} is an assignment of a value to each \Criterion{} in that \ProblemSpace.
\end{definition}

The above is important, because it suggests that, when doing \RE, we can be solving one \RP{} instance, or we may have the freedom to choose the \ProblemInstance{} instance to solve. This choice may be due to necessity, such as when it is not feasible to design the system in such a way, that it maps exactly to the \Criteria{} values chosen in the \ProblemInstance.

For example, in Figure \ref{f:ex:problem-space-ra-var1-var2}, the empty circle is a \ProblemInstance, and if choose to solve it, then we have decided to look for \Specification s which do not satisfy the requirement \req{rA}. If we choose to solve the \ProblemInstance{} marked with the black circle, then we will be looking for \Specification s that satisfy the requirement \req{rA}.

\subsection{\SolutionSpace}\label{s:spaces:solution-space}
The \SolutionSpace{} is made of dimensions that correspond to properties which we can \zi{design into} the system. 

For example, if we have in the \ProblemSpace{} a \Criterion{} that measures the response time of a server (or more abstractly, the responsiveness of the system), then in the \SolutionSpace, we are interested in what we should build into the system, to make sure that it will achieve some value over that \Criterion. 

We introduce the following terms.

\begin{definition}\label{d:solution-space}
\xb{\SolutionSpace:} Set of points, where each coordinate of each point corresponds to the value of a \Parameter.
\end{definition}

\begin{definition}\label{d:parameter}
\xb{\Parameter:} A variable, such that we can choose its value for each \Specification, and this value is expected to influence the behaviour of the system-to-be in some predictable manner.
\end{definition}

We chose the term \Parameter, because its dictionary definition indicates it is a variable whose value \zi{we choose}. This is very different from \Criteria, since the idea for them is that we obtain their values through measurement or otherwise,  rather than set or choose their values.

\begin{figure}[t]
	\centering
	\includegraphics[height=70mm]{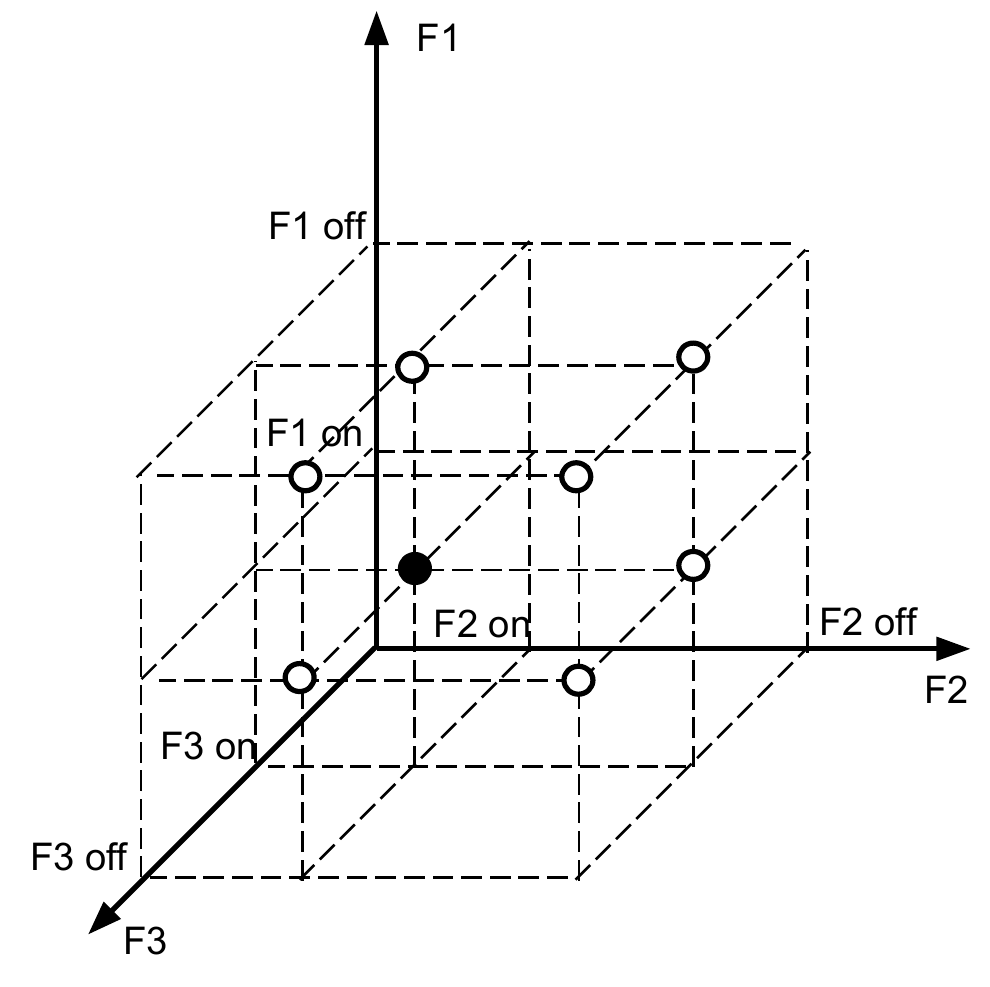}
\caption{\SolutionSpace{} that has 8 points, defined by three \Parameter s, \req{F1}, \req{F2}, and \req{F3}, each of which can either be on (in a \Specification{} of the system-to-be) or off (not in a \Specification). On or off value of one \Parameter{} is assumed independent from the on or off values of other \Parameter s. The black circle is a position where all three are included in the \Specification.}
\label{f:ex:solution-space-8-points}
\end{figure}

\Parameter{} is a general notion, intended to be independent from what one chooses to call the fragments of a \Specification. A two-valued \Parameter{} can capture the common notion of functionality, as something that is either present or absent in the system-to-be. When it can take more values, a \Parameter{} can capture the idea of parameterisable functionalities of the system-to-be; for example, that we \zi{can or need to decide} the resolution of a screen in an operating system, the number of ambulances in a system delivering emergency services, and so on.

As for the \ProblemSpace, there can be relations between values on various \Parameter s in the \SolutionSpace; for example, a functionality \req{F1} may be decomposed into two different functionalities \req{F1a} and \req{F1b}, so that the presence or absence of \req{F1} depends on the presence or absence of both \req{F1a} and \req{F1b}.

\begin{figure}[t]
	\centering
	\includegraphics[height=70mm]{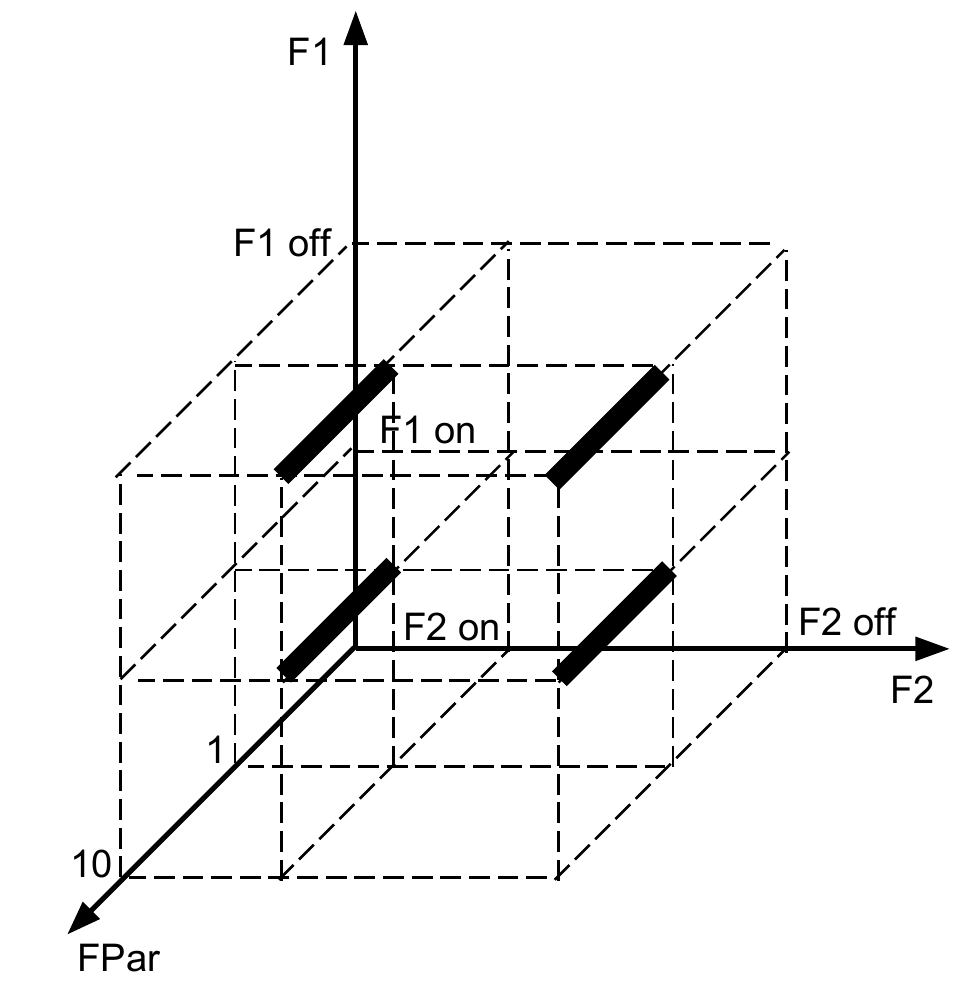}
\caption{\SolutionSpace{} made of all points on the four thick black lines, defined by two two-valued \Parameter s \req{F1} and \req{F2}, and one \Parameter{} \req{FPar}, which can take a real value between 1 and 10.}
\label{f:ex:solution-space-many-points}
\end{figure}

Figure \ref{f:ex:solution-space-8-points} illustrates a \SolutionSpace{} defined by three \Parameter s, each of which can be either included in a \Specification, or excluded from it. Figure \ref{f:ex:solution-space-many-points} illustrates the \SolutionSpace{} generated by two two-valued \Parameter s, and a \Parameter whose value can be any real number between 1 and 10.

\subsection{Double Decision-Making}\label{s:spaces:double-decision-making}
\ProblemSpace{} and \SolutionSpace{} concepts make it clear that \RE{} may involve choosing both the \ProblemInstance{} to solve, and the \Specification{} which solves it. For example, it may be that we change from one \ProblemInstance{} to another because of feasibility, while implementation cost could lead us to change from one \Specification{} to another. 

Problem-solving in \RE{} involves this double and interdependent decision-making, of choosing the \ProblemInstance{} and choosing the \Specification. If we were to design a process for problem-solving, then we would need to decide, for example, if we are going to first choose a \ProblemInstance, and then design the \Specification{} to solve it, or if we would first choose a feasible \Specification, and then see how to make the least changes to it to make sure it satisfies the \ProblemInstance{} closest to the one we should solve. Or if we should do something else, such as distinguish mandatory from nice-to-have values of \Criteria, and choose only among those \Specification s, which satisfy all of the mandatory \Criteria{} values.

For example, the \ZJRP{} gives requirements and domain knowledge, so that we are given the \ProblemInstance, and we need to incrementally design a \Specification, which should be the \Solution{} to exactly that \ProblemInstance. This is because all members of $R$ and all members of $K$ must be satisfied, so that all values of all \Criteria{} in this \ProblemSpace{} are already decided.

This double decision making is an additional argument supporting the idea that \ZJRP{} is not the root of a taxonomy of \RP s. There is no particular reason why we must first choose one \ProblemInstance{} and then look for its \Solution{} in the \SolutionSpace. It can happen that we start from unrealistic requirements, and/or from conflicting requirements, and that, as we proceed in problem-solving, we have to revise the \ProblemInstance{} we are solving -- that is, we need to move in the \ProblemSpace, not only in the \SolutionSpace. And this is the implicit assumption in \RE{} research concerned with requirements inconsistency \cite{hunter1998managing}, conflicts \cite{van1998managing}, and obstacles to requirements satisfaction \cite{van2000handling}.

\subsection{Relating the \ProblemSpace{} and the \SolutionSpace}\label{s:spaces:relationship}
There are two kinds of relations between the \Specification s in the \SolutionSpace{} and the \ProblemInstance s in the \ProblemSpace:
\begin{enumerate}
    \item{The \Solve{} relation, between one \Specification{} and one \ProblemInstance, used to indicate that the former can be a \Solution{} to the latter,}
    \item{The \Depend{} relation, when it makes the values of some \Criteria{} (not necessarily all) depend on values of some \Parameter s.}
\end{enumerate}

\subsubsection{The \Solve{} Relation}\label{s:spaces:relationships:solve}
By being a point in the \SolutionSpace, the \Specification{} can be seen as the synthesis of all our decisions, on what values to give to all \Parameter s. 

Measurement or simulation of a \Specification{} maps it to a Problem Instance in the Problem Space. (Measurement and simulation are not the only ways; there are others, such as relying on expert opinion, but this does matter much in this discussion.) We say that if the \Specification{} X maps to the \ProblemInstance{} Y, then X solves Y. But since there can be many \Specification s that can map to the same \ProblemInstance, we will say that X is the \Solution{} of Y only if it is the one \Specification{} chosen among all others that are considered during problem-solving.

\begin{figure}[t]
	\centering
	\includegraphics[width=110mm]{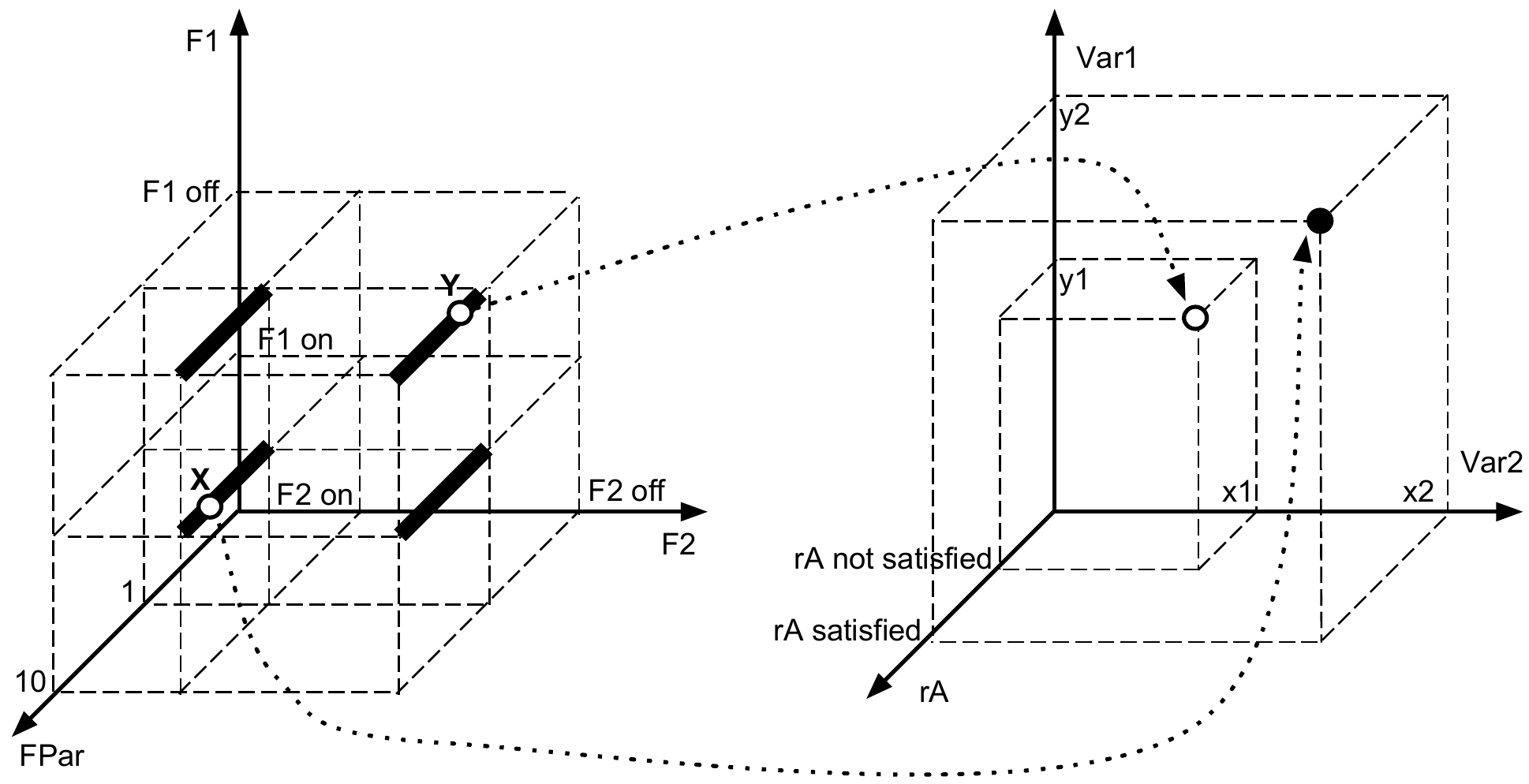}
\caption{Mapping \Specification s from the \SolutionSpace{} (left) to \ProblemInstance s in the \ProblemSpace{} (right). The dotted lines are instances of the \Solve{} relation. Two \Specification s X and Y are marked in the \SolutionSpace. If we are looking to solve the \ProblemInstance{} where \req{rA} is satisfied, then \Specification{} Y will not be appropriate, as it maps to the \ProblemInstance{} in which the requirement \req{rA} fails. The \ProblemSpace{} shown involves non-functional requirements, whose degree of satisfaction is quantified with variables \req{Var1} and \req{Var2}.}
\label{f:ex:problem-space-solution-space-mapping}
\end{figure}

We need a name for the relation between a \Specification{} in the \SolutionSpace{} and a \ProblemInstance{} in the \ProblemSpace. This relation should exist between a \Specification{} and a \ProblemInstance, if we believe that, when that \Specification{} is implemented, measuring it over the \Criteria{} in the \ProblemSpace{} will result in exactly those \Criteria{} values that the \ProblemInstance{} has.

\begin{definition}\label{d:solve}
\xb{\Solve} is a relation from one \Specification{} in the \ProblemSpace{} to one \ProblemInstance{} in the \ProblemSpace. We say that \Specification{} A \Solve s \ProblemInstance{} B if and only if we believe that, if the system is made according to \Specification{} A, and we measure the system according to the \Criteria{} that define the \ProblemSpace{}, then we will obtain values of these \Criteria which correspond to the values that they have in \ProblemInstance{} B.
\end{definition}

For a given \RP, instances of the \Solve{} relation are the result of problem-solving, as they can be found only after at least one \Specification{} and one \ProblemInstance{} have been identified.

\subsubsection{The \Depend{} Relation}\label{s:spaces:relationships:depend}
The \Depend{} relation is between \Parameter s and/or \Criteria, when their values are interdependent. It is not restricted to being between individual points in the \ProblemSpace{} and the \SolutionSpace. It can be used to represent that, for example, the value of a \Parameter{} depends on the values of other \Parameter s and/or \Criteria, that the value of a \Criterion{} depends on values of other \Criteria{} and/or \Parameter s, that the value of a \Parameter{} has to be in some range, and so on.

A simple way to think about the \Depend{} relation, is that, any function that relates the values of \Parameter s and/or \Criteria{} is an instance of the \Depend{} relation.

\begin{definition}\label{d:depend}
\xb{\Depend:} Given some variables, which may be \Criteria{} and/or \Parameter s, if their values are interdependent, then there is an instance of the \Depend{} relation between these variables.
\end{definition}

\begin{figure}[t]
	\centering
	\includegraphics[width=110mm]{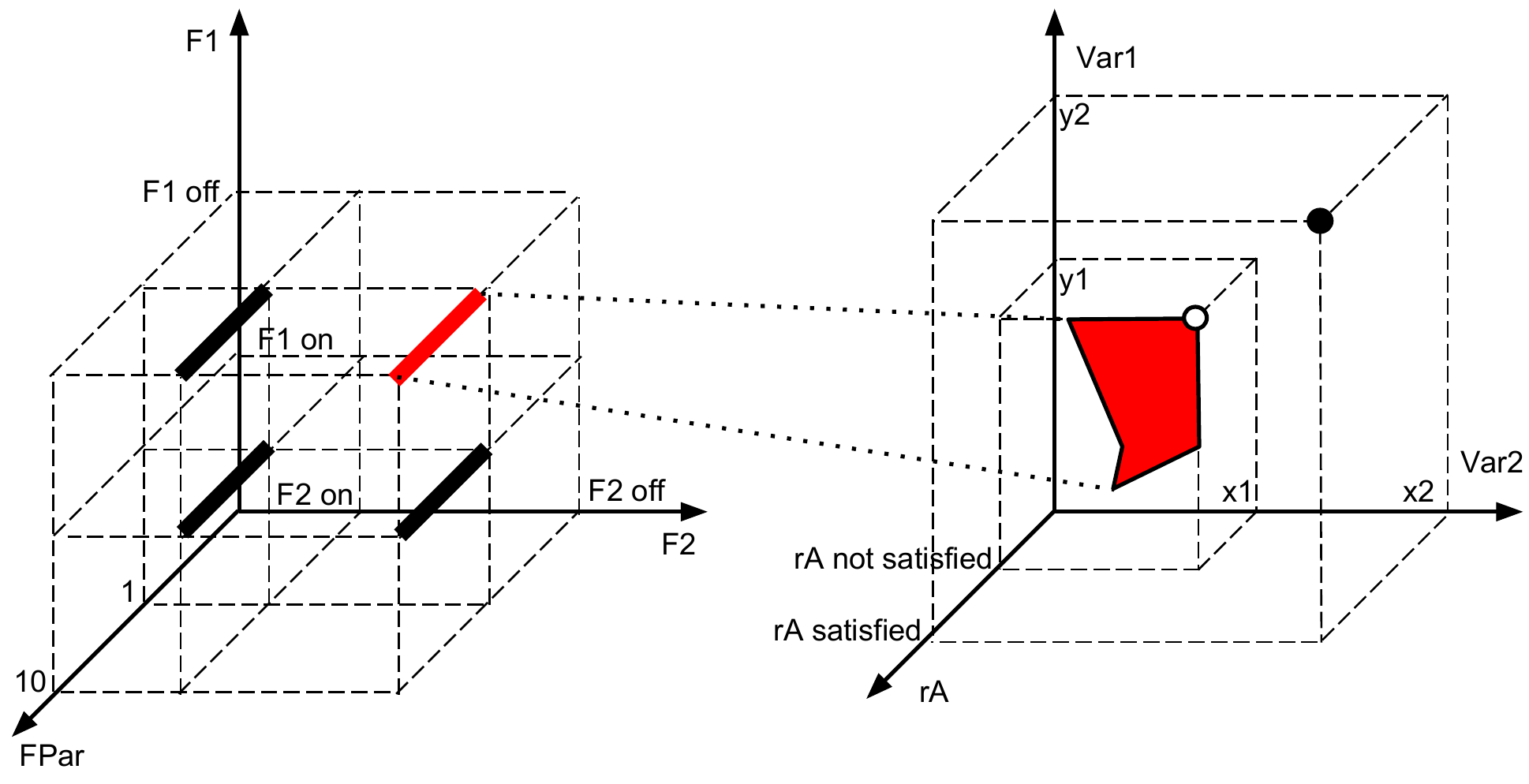}
\caption{If a point is chosen along the highlighted line in the \SolutionSpace{} (left), then this results in a point on the highlighted area in the \ProblemSpace{} (right). The figure illustrates the \Depend{} relation between the values of the \Parameter s \req{Fpar}, \req{F1}, and \req{F2}, and the values of \Criteria{} \req{rA}, \req{Var1}, and \req{Var2}.}
\label{f:ex:depend-relation}
\end{figure}

Figure \ref{f:ex:depend-relation} gives an illustration of a function where the value of \Criteria{} depend on the value of \Parameter s.

\section{\OptimalSpecification s in \ProblemSpace s}\label{s:optimal-specifications}
Our definition of the \Solution{} concept in Section \ref{s:spaces} only says that the \Solution{} is that \Specification{} which we commit to, as the \Specification{} which the system-to-be should implement. 

In contrast, the discussion of optimality in Section \ref{s:uniqueness} used the obvious premise that it is desirable, in general, to commit to the \Specification{} which is somehow the ``best'' relative to those \Specification s that are considered during problem-solving. 

The goal now is to relate these two ideas, of optimality of, and commitment to a \Specification, and do so using the concepts and relations introduced so far. 

Relating commitment and optimality means using the following revised \Solution{} definition.

\begin{definition}\label{d:solution:revised}
\xb{\Solution{} (replaces Definition \ref{d:solution}):} In a given \RP, and thereby for its \ProblemSpace{} and \SolutionSpace, the \Solution{} is the \OptimalSpecification.
\end{definition}

The revision reflects the idea that relating commitment to optimality amounts to asking that we commit to the \Specification{} which satisfies those properties that optimality imposes. This is different from the original \Solution{} in Definition \ref{d:solution}, where the only property of the chosen \Specification{} is that we commit to it, regardless of how exactly it relates to other \Specification s.

\subsection{Preference and Utility in \ProblemSpace s}\label{s:optimal-specifications:preference-utility}
The revised \Solution{} concept leads to the question of when a \Specification{} is also an \OptimalSpecification, in the terminology of \ProblemSpace s and \SolutionSpace s.

To answer this, recall that \Specification s in themselves are interesting only because they describe systems, which, \zi{if} they are implemented accordingly, would achieve specific values over all of the \Criteria{} in the \ProblemSpace.\footnote{As we cannot make all systems that implement all the \Specification s, and then measure values of \Criteria{} on systems themselves, we make the simplifying assumption that it is a \Specification{} that maps to points in the \ProblemSpace, not the system; this changes nothing in this discussion, other than pointing out that the mappings between the \SolutionSpace{} and \ProblemSpace{} will often simply be based on our experience, predictions, speculation, and such, not on actual measurement.} It is because it manages to map to some \Criteria{} values that a \Specification, or a set of \Specification s is of any relevance in problem-solving.

The fact that we are designing \Specification s because we are in fact interested in some values of \Criteria, means that \zi{whether a \Specification{} is better than another is an issue that is solved not by looking only at the \SolutionSpace, but at where a \Specification{} maps to in the \ProblemSpace}. 

Therefore, whether a \Specification{} is the \OptimalSpecification{} depends on \zi{where it maps in the \ProblemSpace}, since different \Criteria{} values are not equally desirable. This was illustrated with indifference curves in Figure \ref{f:ex:indifference-curves} earlier: we may be indifferent between some value combinations, which we can represent with an indifference curve, while different indifference curves reflect that we can evaluate some value combinations as strictly more desirable than others.

In the ideal and almost certainly infeasible case, we would have information about preference between every combination of values of independent \Criteria. This would give us the indifference curves for these combinations, and also the corresponding utility function. In more realistic cases, we may be able to find a partial order preference relation, which compares some combinations, perhaps over a subset of \Criteria{} in the \ProblemSpace.

In any case, however, the important observation to make is that \zi{in order to say which regions of the \ProblemSpace{} are more desirable than others, it is necessary to have information about the relative desirability, that is, of preference of \Criteria{} value combinations}. 

As a brief digression, we recall here how the notions of \zi{preference}, \zi{utility}, and \zi{indifference curves} are related in general, in economics. The term preference is used here to mean preference relation, the binary relation that indicates the relative desirability between two things; in this paper, it is the binary relation that compares the relative desirability of \Criteria{} values, be they values of the same \Criterion, or of different \Criteria, or of value combinations of \Criteria. Utility is a quantitative representation of information in a preference relation. If a preference relation, for example, over different values of a single \Criterion, is transitive, complete, and continuous, then we can define a corresponding utility function which is continuous. An indifference curve is a set of different combinations of things compared in terms of preference, which are all equally preferred. A utility function reflecting that preference relation will return the same utility value for all members of that set. The same utility function gives many indifference curves, as there can be several sets of equally preferred combinations, such that combinations from different such sets are not equally preferred. As a final remark, utility is a generic notion, which can have different interpretations, and the specific meaning it will have depends on what one chooses to quantify desirability with.

\subsection{Utility Representation with \Criteria{} and \Depend{} Relations}\label{s:optimal-specifications:utility-representation}
\begin{figure}[t]
	\centering
	\includegraphics[width=110mm]{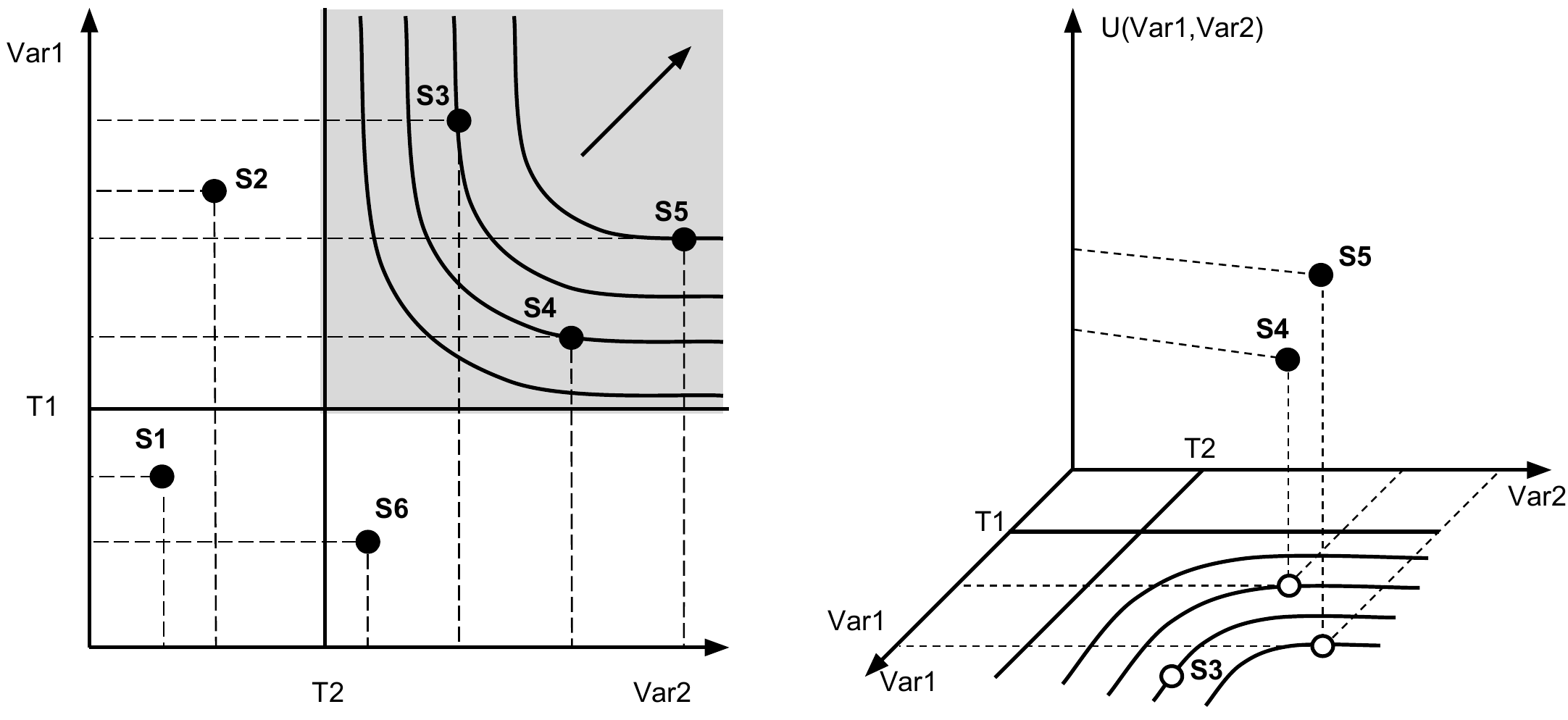}
\caption{The figure on the left shows a part of the \SolutionSpace, borrowed from Figure \ref{f:ex:indifference-curves}. S3, S4, and S5 indicate mappings of three different \Specification s to this part of the \ProblemSpace. The three are above the threshold values \req{T1} and \req{T2}. Each is on a different indifference curve, indicating that they are not equally preferred. The part on the right adds a \Criterion{} \req{U(Var1,Var2)} which gives the utility value of value combinations of \req{Var1} and \req{Var2}. Note that the figure on the right is a simplification, as the two points S4 and S5 are points on a three-dimensional surface; the shape of the indifference curve hints at the shape of that surface.} 
\label{f:ex:indifference-curves-and-utility}
\end{figure}

This preference information can be represented using the notions which were already introduced, namely, \Criteria{} and the \Depend{} relation. To add a utility function, add a \Criterion{} whose values are read as utility values, and a \Depend{} relation that specifies which other \Criteria{} values determine the utility value. This is the illustrated in Figure \ref{f:ex:indifference-curves-and-utility}, by adding a utility \Criterion{} \req{U(Var1,Var2)} which gives the utility value for combinations of \req{Var1} and \req{Var2} \Criteria.

It is important to note that there can be different utility \Criteria{} in the same \ProblemSpace. This can be due to the fact that we may know preferences between value combinations of some \Criteria, but not of others; or it may reflect differences in preferences between stakeholders, in which case we could have different utility \Criteria{} for different stakeholders. Another important remark is that any definition of the dependence between the value of some \Criteria{} and the value of other \Criteria, be they utility \Criteria{} or otherwise, are defined by a function that consequently is an instance of the \Depend{} relation.

\subsection{\DecisionRule s}\label{s:optimal-specifications:decision-rules}
Even in the trivial example introduced in Section \ref{s:optimality-criteria-importance:switching}, we can have many different preference orders, and thereby different utility functions. For example, there can be a preference order over values of \req{Var1}, another over those of \req{Var2}; if there are three stakeholders, A, B, and C, and they all have their own preferences over the values of these \Criteria, then we might have nine preference orders in total; worse, we may not have a rule that tells us how to obtain preferences over combinations of \req{Var1} and \req{Var2}, from those that we have, over only values of \req{Var1} and only values of \req{Var2}.

As soon as there are two preference relations, it is necessary to explain how they are used \zi{together} to compare \ProblemInstance s to which \Specification s map in the \ProblemSpace. 

This explanation is called the \zi{\DecisionRule}, and can take various forms. 

In the example illustrated in Figure \ref{f:ex:indifference-curves-and-utility}, we may want to do one of the following:
\begin{itemize}
    \item{Maximize the value of \req{U(Var1, Var2)},}
    \item{Maximize the value of \req{Var1},}
    \item{Maximize the value of \req{Var2},}
    \item{And so on.}
\end{itemize}

Each item gives the condition that a \Specification{} should satisfy, in order to be the \OptimalSpecification. In the first item above, the \DecisionRule{} means that the \OptimalSpecification{} is the \Specification{} which maximizes the value of \req{U(Var1, Var2)}. In the second item, notice that the utility function is not given, but is implicit: namely, utility is independent from the value of \req{Val2}, and increases with the increase in \req{Val1}. The third item is the opposite case. 

\begin{definition}\label{d:decision-rule}
\xb{\DecisionRule:} A \DecisionRule{} is the \Criterion, such that in a given set of \Specification s, the \Specification which has the highest value on that \Criterion, is the \OptimalSpecification.
\end{definition}

The \DecisionRule{} is a special \Criterion, as it will single out the \OptimalSpecification{} among those \Specification s that we are considering. Its value may, and often will be the result of combining values of other \Criteria, such as when \req{U(Var1,Var2)} is the \DecisionRule.

\section{\ROPfull s}\label{s:rop}
The \DecisionRule{} definition, together with the notions we had introduced so far, introduces a new class of \RP s, all of which include the \DecisionRule{} defined above. They are called \ROPfull s (\ROP s). 

Because they all include the \DecisionRule, they are not a subclass of the \ZJRP.

\ROP s are important for the \RPAS, as we will argue in Section \ref{s:rpas} that the \RPAS{} amounts to  a set of \ROP s, together with some additional constraints on that set.

This section first defines the \ROP, and then illustrates how the \ZJRP, if we only apply one change to it, becomes a subclass of the \ROP.

\subsection{Problem Statements}\label{s:rop:definition}
Solving an \ROP{} involves doing design and doing decision-making. 

Here, \zi{doing design} means doing four tasks, and not necessarily in the sequence given below: 
\begin{enumerate}
    \item{Constructing the \ProblemSpace, by finding and defining \Criteria, and defining \Depend{} relations between these \Criteria, when we know that there are correlations between their values, or have other reasons to believe that their values determine one another.}
    \item{Constructing the \SolutionSpace, by identifying and defining \Parameter s, and the \Depend{} relations between the \Parameter s.}
    \item{Defining \Depend{} relations between \Parameter s and \Criteria, so as to clarify how the choices of values of the former relate to the satisfaction of the \Criteria.}
    \item{Choosing the \DecisionRule, and defining the \Depend{} relation which makes its value depend on values of other \Criteria{} in the \ProblemSpace.}
\end{enumerate}

\zi{Doing decision-making} here means identifying and committing to values of \Parameter s which, relative to any other combination of \Parameter{} values, result in the highest value of the  \DecisionRule. 

This separation between design and decision-making is introduced to highlight that there are two kinds of tasks in \ROP{} problem-solving. The aim is \zi{not} to suggest that they are necessarily done in sequence, that we first have to do all related to design, and then do the decision-making. The separation still fits incremental design, which, in this perspective, amounts to a sequence of activities, some of which are focused on design (such as, adding new \Criteria{} and \Parameter s, choosing values thereof, etc.) and some of which are concerned with decision-making (trying to find values of \Parameter s) and can initiate the next design iteration (such as if we fail to find the \Parameter{} values, which may lead to changing \Parameter s, \Criteria, \Depend{} relations, and so on).

The two problems are defined as follows.

\begin{definition}\label{d:rdp}
\xb{\RDPfull{} (\RDP):} \xb{Given} the information about the stakeholders' expectations, and the information about the environment of the system-to-be, \xb{define} (i) the \ProblemSpace, (ii) the \SolutionSpace, (iii) the \Depend{} relations over \Criteria{} in the \ProblemSpace{} and the \Parameter s in the \SolutionSpace, and (iv) the \DecisionRule.
\end{definition}

\begin{definition}\label{d:rop}
\xb{\ROPfull{} (\ROP):} \xb{Given} (i) a \ProblemSpace, (ii) a \SolutionSpace, (iii) a set of \Depend{} relations between \Parameter s and/or \Criteria, (iv) a \DecisionRule, and (v) a set of \Parameter s in the \SolutionSpace{} whose values have not been set, called the \DecisionSet{}, \xb{find} the values of \Parameter s in the \DecisionSet, such that there is no other combination of values of these same \Parameter s, which returns a higher value of the \DecisionRule.
\end{definition}

As \Criteria{} and \Parameter s are variables, and \Depend{} relations are functions over these variables, the \ROP{} can be rewritten as follows:
\begin{eqnarray*}
& & \zi{Maximise }\; f_{0}(x) \\
& & \zi{subject to }\; f_{i}(x) = b_{i}, \zi{ for } i = 1, \ldots, n
\end{eqnarray*}

where:
\begin{itemize}
    \item{$x$ is the \DecisionSet, that is, a set of \Parameter s in the \SolutionSpace, whose values should be set,}
    \item{$f_{0}(x)$ is the \Depend{} relation that returns the value of the \DecisionRule{} in the \ProblemSpace,}
    \item{Each $f_{i}(x) = b_{i}$, for $i = 1, \dots, n$ is a \Depend{} relation instance.}
\end{itemize}

\subsection{An Illustration: Converting the \ZJRP{} into an \ROP{} Subclass}\label{s:rop:zjrop}
The aim in this section is to illustrate how \ROP s are related to the \ZJRP. We do this by rewriting the \ZJRP{} as a \ROP{} subclass. That subclass will be called \ZJROPfull{} (\ZJROP).

Defining \ZJROP{} involves answering the following questions:
\begin{enumerate}
    \item{What are $K$, $R$, $S$ of the \ZJRP{} in the \ROP?}
    \item{What are the \Depend{} relations in \ZJRP?}
    \item{What is the \DecisionSet{} in \ZJRP?}
    \item{What is the \DecisionRule{} in \ZJRP?}
\end{enumerate}

The \ZJRP{} uses the syntactic consequence relation $\vdash$, leading us to assume that it is of classical, two-valued logic, so that members of $K$, $S$, and $R$ can either take the value 0 or 1. Note that these values have a different reading depending on them being for members of $K$, $S$, or $R$: in $R$, ``1'' tends to read ``satisfied'', ``achieved'', etc.; in $K$, ``1'' can read ``maintained'', ``satisfied'', etc.; and in $S$, ``1'' can read ``implemented'', ``configured'', or otherwise, along these lines. 

It follows that the \ZJROP{} has only integer binary variables. 

In the terminology of \ProblemSpace s and \SolutionSpace s, variables that correspond to members of $K$ and $R$ are \Criteria{} in \ZJROP, and define the \ProblemSpace; variables that correspond to members of $S$ are \Parameter s, and define the \SolutionSpace.

The \Depend{} relations in \ZJROP{} correspond to the relations between the members of $K$, $S$, and $R$. For example, if the value of a variable $w_{i}$ depends on the value of other variables $y_{1}, \ldots, u_{k}$, then we define a \Depend{} relation $w_{i} - f(y_{1}, \ldots, u_{k}) = 0$. 

We will have such \Depend{} relations, because if a requirement \req{rA} is, for example, refined by two other requirements \req{rA1} and \req{rA2}, then \req{rA} is satisfied if and only if both \req{rA1} and \req{rA2} are satisfied. In \ZJROP, this is captured by a \Depend{} relation.

In the \ZJRP, $R$ is given and must be satisfied, which in the \ZJROP{} means that the values of all variables from $R$ are set to 1, and they therefore cannot be in the \DecisionSet. $K$ is also given, and since $R$ and $S$ should be consistent with it, we can also set all $K$ variables to 1. 

Variables from $S$ in the \ZJRP{} remain as the members of the \DecisionSet{} of the \ZJROP. If an $S$ variable depends on the values of other $S$, $K$, or $R$ variables, then we will exclude it from the \DecisionSet. It follows that the cardinality of the \DecisionSet{} does not need to equal cardinality of $S$.

All members of $K$ and $R$ are equally preferred, as it is equally important to satisfy every member of $R$, and to maintain every member of $K$. We concluded earlier that there are no means to compare specifications in the \ZJRP. There is no information about preferences in the \ZJRP.

But once we allow alternative refinement of same requirements, we have alternative \Specification s to choose from, and we want the \ZJROP{} to reflect this. Suppose that we allow any member of $K$, $R$, $S$ to be refined. This means that refining $R$ gives us another set of requirements, which includes requirements that refine those in $R$. Let that set be $\zi{ref}(R)$, and let $\zi{ref}(K)$ and $\zi{ref}(S)$ be for $K$ and $S$, what $\zi{ref}(R)$ is for $R$.

The resulting \ZJROP{} is to find a subset of $\zi{ref}(S)$, such that, if all \Parameter s in that subset are set to 1, then all \Criteria{} from $K$ and $R$ obtain the value 1. As there are is no preference information in the \ZJRP, it is unclear what the \DecisionRule{} should be.

The simplest choice we saw for the \DecisionRule{} in the \ZJROP, is that it is equal to the sum of the members of $\zi{ref}(S)$, meaning that we want to find the smallest subset of $\zi{ref}(S)$ which manages to result in the assignment of the value 1 to all \Criteria{} corresponding to members of $K$ and $R$. So we want to maximise the following:

\begin{eqnarray*}
    & & -\sum_{x_{i} \in \zi{ref}(S)}x_{i}
\end{eqnarray*}

In addition to the \Depend{} relations for refinements, we need the following \Depend{} relations to guarantee that the \Solution{} must assign the value 1 to every \Criterion{} from $K$ and $R$:

\begin{eqnarray*}
    & & \sum_{y_{j} \in R}y_{j} = |R| \\
    & & \sum_{z_{l} \in K}z_{l} = |K|
\end{eqnarray*}

Note that, since every $x_{i}$ in $\zi{ref}(S)$ is a binary variable, the solution to \ZJROP{} will be the smallest subset of $\zi{ref}(S)$, which satisfies all \Depend{} relations. This modifies the \PropDR{} from the \ZJRP; that property is neutral about the specifics of $S$, as long as it satisfies the \PrConsistency{} and \PrSatisfaction{} properties. This makes the \ZJROP{} a different problem than \ZJRP, and this is not necessarily a relevant problem: the number of \Parameter s set to 1 in $\zi{ref}(S)$ has, in itself, nothing to do with the quality expected from the system-to-be.

\section{The \RPASfull s}\label{s:rpas}
This section introduces the definition of the \RPAS{} in the following steps. In the first step, in Section \ref{s:rpas:key-ideas}, we recall the key ideas in \RE{} for \AS s. In Section \ref{s:rpas:premises}, we relate these ideas to the terminology of \ProblemSpace s and \SolutionSpace s, the \RDP, and the \ROP. This leads us in Section \ref{s:rpas:statements} to the definition of the design and decision-making problems that form the \RPAS.

\subsection{Key Ideas in \REfull{} for \ASfull s}\label{s:rpas:key-ideas}
In order to adapt to changes, the \ASfull{} has to be capable of detecting changes; it can only respond and adapt to those changes that it can detect.

To detect changes, the \ASfull{} has to gather data about events in its operating environment and about the functioning of its own components. At all times, and on the basis of these observations, the \ASfull{} has to estimate the level to which it satisfies the stakeholders' requirements. If the levels of satisfaction are inadequate, the \ASfull{} has to make changes to what it does in order to satisfy the requirements. 

The \AS{} changes its behaviour via \zi{feedback loops}. A feedback loop defines the variables whose values need to be monitored; the values would be collected by sensors, or would be functions of variables whose values the sensors collect. When the values fall out of the predefined and allowed range, this triggers functionality in the \AS, dedicated to make changes to the operation of the \AS.

Capability to adapt requires a hierarchy of functionality in an \AS. The lowest-level functionality interacts with the environment; the next level is functionality that enables feedback loops, which monitor signals from sensors that monitor the environment and the functionality at the lowest level; the second level is functionality that enables feedback loops that monitor and manipulate the feedback loops at the first level; and so on.

The above leads to key observations about the \zi{run-time} of \ASfull. (i) The level at which requirements are satisfied will vary, due to failure in the system and change in its environment. (ii) It is necessary to monitor the level of satisfaction of requirements, in order to know when the system needs to adapt. (iii) When the system adapts, it may have different ways of adapting, and each of these ways may have a different impact on requirements satisfaction levels. (iv) Whenever it needs to adapt, the system should adapt in the way that optimizes the levels of requirements satisfaction, relative to the newly observed failure of a component, or of a change in the environment.

The observations about run-time have important implications for the \zi{design-time} of \ASfull s.

Due to observation (i), it may be too idealistic and impractical to think of requirements as being either satisfied or not, since this may lead to too many failed requirements, too often. It can be more practical, therefore, to define multivalued scales of requirements satisfaction, where failure equates to only some of many values. This is done through the \xb{relaxation of requirements} \cite{letier2004reasoning, whittle2009relax, baresi2010fuzzy}, where the idea is to replace binary levels of satisfaction with, for example, continuous scales of satisfaction, or by letting the requirement be binary, but tracking the frequency of them being satisfied or failing, then using that frequency as the measure of the degree to which these requirements are satisfied. 

Observation (ii) suggests that it is necessary, at design-time, to define the levels of requirements satisfaction that trigger adaptation. If the requirement has a binary satisfaction scale, being either satisfied or not, it may be relevant to define the minimal probability of observing its satisfaction; for example, in an ambulance dispatch system, asking for the probability of at least 0.95, that an ambulance arrives to an incident location within 5 minutes of being dispatched to it. This would translate, at run-time, into looking at the frequency of incidents where the ambulance arrived 5 minutes or more, and triggering adaptation if that frequency is 5\% or more of all incidents to which an ambulance was dispatched. If the requirement has a scale with many levels of satisfaction, then a threshold value has to be defined on that scale, such that, when the satisfaction is below threshold, the system needs to adapt. This has led to research on \xb{awareness requirements} \cite{souza-et-al:sesas13}, which are used to define these thresholds for triggering adaptation.

At run-time, when awareness requirements are satisfied, feedback loops become active, and the system adapts. Because of observations (iii) and (iv), it is necessary to define at design-time the requirements that the system should satisfy when adapting. These are the so-called \xb{evolution requirements} \cite{Souza+:2012:CSRD}, and place constraints on how the system adapts. In the terminology of research on the \RE{} for \ASfull s, evolution requirements place constraints on the range of \xb{reconciliation tactics} \cite{Fickas+:1995:RE,Feather+:1998:IWSSD,Robinson:2006:REJ} that the system may choose to apply, when adapting. The ambulance dispatch system could adapt to the failure in its component that records ambulance location, by requiring that control assistants who dispatch ambulances, record these locations manually, or by relying on the record of ambulance location by the part of the system which is located in each ambulance. An evolution requirement may indicate that control assistants should not manually record information, unless more than one of the automatic data recording components fails; this would exclude the second adaptation in the example.

\subsection{Premises for the definition of the \RPAS}\label{s:rpas:premises}
The aim in this section is to relate the key ideas from existing research, to the terminology of \ProblemSpace s, \SolutionSpace s, the \RDP, and the \ROP. 

\subsubsection{Monitoring}\label{s:rpas:monitoring}
The ability of an \AS{} to adapt to changes requires that the system can detect changes. We introduce two terms, in order to talk about the extent of changes that the \AS{} is designed to detect.

\begin{definition}\label{d:monitored-variable}
A \xb{\MonitoredVariable{}} is a variable whose values the \ASfull{} collects and whose changes of values can trigger adaptation.
\end{definition}

\begin{definition}\label{d:monitoring-scope}
The \xb{\MonitoringScope} of an \AS{} is the set of all of its \MonitoredVariable s and, for each variable, the range of values that the \ASfull{} can detect.
\end{definition}

The \MonitoringScope{} describes what the \AS{} is able to detect as change in its environment, and change in its own functionality. These changes are detected via sensors. The variety and the specifics of the sensors that an \AS{} has, determine its \MonitoringScope.

If something in the environment, or in the \AS{} itself changes, but there are no \MonitoredVariable s to reflect that change, then the \AS{} will ignore it. 

The \MonitoringScope{} reflects the \zi{changes that were anticipated at design-time}. All other changes, which the \MonitoringScope{} cannot detect, remain as \zi{unanticipated} changes. It is in this sense that we talk about \zi{scope} in \MonitoringScope, as the scope of changes that have been predicted and considered as particularly relevant at design-time, regardless of how relevant they may actually prove to be at run-time.

The design of the \MonitoringScope{} involves choosing the \MonitoredVariable s, based on the sensors that can be built into the \AS, and the \Depend{} relations between \MonitoredVariable s and the \Parameter s in the \SolutionSpace, and the \Criteria{} in the \ProblemSpace. Without these \Depend{} relations, it is unclear why sensors would be used at all, or why and when the \AS{} should adapt. 

Monitoring the level of requirements satisfaction here means having \MonitoredVariable s that are equal to some of the \Criteria{} in the \ProblemSpace. Monitoring the satisfaction of domain knowledge also means that the \MonitoringScope{} will include \MonitoredVariable s that are equivalent to some \Criteria, when these \Criteria{} reflect domain knowledge. We may also have \MonitoredVariable s that monitor \Parameter{} values, as we want to know when failure happens, that is, when actual \Parameter{} values are not those defined in the \Solution.

\subsubsection{Change}\label{s:rpas:premises:change}
The \MonitoringScope{} is unlikely, in general, to be such that it enables the \AS{} to detect all relevant changes in its functioning, its environment, and in the expectations of its stakeholders. This is the case as we cannot, at design-time, anticipate all that could potentially change, and to which the system should adapt.

Independently from the \MonitoringScope, there is what we call the \zi{\ChangeScope}, denoting the variety of phenomena in, or outside the \AS, which may occur, and to which the \AS{} \zi{may} need to adapt.

The \ChangeScope{} is not limited to phenomena that can cause the \AS{} to fail to satisfy its design-time requirements, and/or to phenomena that put it at odds with its environment. 

Stakeholders need to perceive the services that the \AS{} delivers as being of high quality. People's evaluations of service quality reflect their own comparisons between expectations and experience with the service \cite{parasuraman1985conceptual, zeithaml1990delivering, pitt1995service, zeithaml1996behavioral, kettinger2005zones}. While design-time requirements are fixed, we may well have an \AS{} that does achieve these requirements, but is perceived as being of low quality; stakeholders' expectations may have changed, enlarging the gap between what they expect, and their experience of the \AS. The following definition of the \ChangeScope{} reflects this.

\begin{definition}\label{d:change-scope}
The \xb{\ChangeScope{}} is the set of variables that describe the phenomena in the environment of the \AS, and/or the system itself, are such that the values of these variables can change independently from the system's operation, and these changes influence the stakeholders' perception of the quality of the \AS.
\end{definition}

At design-time and run-time, then, there can be many variables in the \ChangeScope{} that the system would ideally monitor. Their relevance may become apparent only after some phenomena occur at run-time and affect the stakeholders' perception of quality of the \AS. In such cases, the engineers need to determine how to measure these phenomena, which sensors to use to collect measurements, and thereby add new variables to the \MonitoringScope.

In the terminology of the \MonitoringScope{} and the \ChangeScope, the design of feedback loops can be described as the task of identifying phenomena that can influence stakeholders' expectations, finding ways to measure them, adding these variables to the \ChangeScope. Next, it is necessary to determine how these variables are related to \Criteria{} and \Parameter s. All such variables are candidates for becoming \MonitoredVariable s. 

It is unlikely that we can identify every variable in the \ChangeScope. Of those that we do manage to identify, we may also be able to make only some into \MonitoredVariable s, due to, for example, there being no sensors that are capable to capture their values.

The conclusion that is important for the definition of the \RPAS, is that the \zi{design} part of the \RPAS{} involves \zi{finding and choosing variables in the \ChangeScope, that need to be made into \MonitoredVariable s in the \MonitoringScope{}}, as it is unlikely that we can ensure that the \MonitoringScope{} fully covers the \ChangeScope. 

\subsubsection{Stability and Adaptation}\label{s:rpas:premises:adaptation}
We will use the term \zi{event} to refer to any change of values of variables in the \ChangeScope; an event can be the result of a failure of a component of the \AS, a drop in the level to which the \AS{} satisfies a requirement, a change in the conditions in the operating environment of the \AS, and so on.

The reason for having \MonitoredVariable s in the first place, is because their changes of values result in changes of values of \Criteria{} in the \ProblemSpace{} and/or \Parameter s in the \SolutionSpace.

The run-time of the \AS{} is, then, a sequence of two kinds of time periods, called periods of \zi{stability} and of \zi{adaptation}. 

Stability is any time period during the run-time of an \AS, during which one of the following conditions holds:
\begin{itemize}
    \item{Values of \ChangeScope{} are not changing;}
    \item{Values of \ChangeScope{} variables are changing, but these variables are not \MonitoredVariable s;}
    \item{Values of \ChangeScope{} variables are changing, and some or all of them are \MonitoredVariable s. The changes result in new values for \Criteria{} and/or \Parameter s. However, these changes are within some ranges that we judged tolerable at design-time.}
\end{itemize}


As the values of \MonitoredVariable s influence values of \Criteria, they also influence the \ProblemInstance{} that the \AS{} needs to be solving. As the values of \MonitoredVariable s change, so does the \ProblemInstance{} to solve. At some time at the run-time, the \AS{} should run according to the \OptimalSpecification{} that solves the \ProblemInstance{} applying at that time period. 

Adaptation is the time period during the run-time of an \AS, when the \AS{} is running according to the \Specification{} which is \zi{not} the \OptimalSpecification{} for the \ProblemInstance{} that the \AS{} should solve at that time. 

To clarify this, suppose that $T_{1}$ denotes a time period, during which the \MonitoredVariable s result in \Criteria{} values which give one \ProblemInstance{} $A$, and the \OptimalSpecification{} $X$ solves $A$. During $T_{1}$, the \AS{} runs according to $X$. 

Let $T_{2}$ be a time period which immediately follows $T_{1}$. $T_{2}$ starts with values of \MonitoredVariable s that give the \ProblemInstance{} $B$. If $B \neq A$, and $X$ does not solve $B$, it is necessary to find a new \Specification{} $Y$ which is the \OptimalSpecification{} for $B$. 

Adaptation is the period during which the \AS{} is searching for this new \Specification, and changes behavior from running according to $X$, to run according to $Y$.

\subsubsection{Relaxation}\label{s:rpas:premises:relaxation}
If the \AS{} were to react to every change in \Criteria{} and \Parameter{} values, stability periods would be shorter, and more resources would be invested in adaptation. 

Relaxation is used to allow the \AS{} to run according to the same \Specification{} in a broader range of conditions, and thereby potentially lengthen stability periods. We have three possible cases, depending on what changes for the \AS, and relaxation can work in each case:
\begin{itemize}
    \item{If only \Criteria{} values change, then the goal is no longer to solve the \ProblemInstance, denote it $A$, that was relevant in the last stability period, but a new \ProblemInstance, denote it $B$.
If the \Specification{} $X$ was the \Solution{} to $A$, and it is not the \OptimalSpecification{} for $B$, then adaptation would involve finding the \Specification{} $Y$, which is the \OptimalSpecification, and thereby the \Solution{} to $B$. Adaptation can be avoided, if we allow $X$ to be the \Solution{} to both $A$ and $B$.}
    \item{If only \Parameter{} values change, then the \ProblemInstance{} $A$ to solve has not changed, but the \AS{} is no longer running according to some \Specification{} $X$ from the last stability period, but according to a new \Specification{} $Y$. There is no guarantee that $Y$ is the \OptimalSpecification{} for $A$, but it can be the \OptimalSpecification{} for some other \ProblemInstance{} $B$. Adaptation would amount to finding a third \Specification, which solves a \ProblemInstance{} $C$, whereby $C$ is closer to $A$ than $B$. To avoid adaptation in this case, relaxation would consist of allowing any one \Specification{} from some set of \Specification s to be the \Solution{} to $A$.}
    \item{If both \Criteria{} and \Parameter{} values are changing, then adaptation will involve finding the \OptimalSpecification{} to the new \ProblemInstance. To avoid adaptation, relaxation would need to be such that, the new \ProblemInstance{} is considered sufficiently close to the one from the last stability period, and the new \Specification{} as the \Solution{} to the new \ProblemInstance.}
\end{itemize}

In the cases above, and using the terminology of \ROP s, adaptation amounts to the act of recomputing the \Solution{} to a \ROP, when the \ProblemInstance{} changes. Relaxation, in contrast, is the act of not triggering the computation of a new \Solution{} to a new \ProblemInstance. At design-time, relaxation consists of changing the \Criteria{} in the \ProblemSpace, and changing \Depend{} relations that link \Parameter{} values to \Criteria{} values. For example, suppose that we have formulated a \ROP, and we want to relax it so as to allow more \Specification s to be its \Solution.


%
\subsection{Problem Statements}\label{s:rpas:statements}
\RPAS{} is a double problem, one focused on design, the other on decision-making. They are defined as follows.

\begin{definition}\label{d:rdpas}
\xb{\RDPASfull s (\RDPAS):} \xb{Given} the information about the stakeholders' expectations, the information about the environment of the system-to-be, and the predictions of changes to stakeholders' expectations and the environment, \xb{define} (i) the sequence of \ROP s that the \AS{} is expected to solve, and (ii) the \MonitoringScope{} needed to detect some or all of the predicted changes. 
\end{definition}

\begin{definition}\label{d:ropas}
\xb{\ROPASfull s (\ROPAS):} \xb{Given} a sequence of \ROP s that the \AS{} is expected to solve and the \MonitoringScope{} for the \AS, \xb{find} the set of \Specification s and \EvolutionRequirement s, such that, if the \AS{} can run according to the \Specification s, and while adapting satisfy the \EvolutionRequirement s, then it will maximise the time that it runs according to the \OptimalSpecification{} in each stability period.
\end{definition}

\section{\ROP{} and Mathematical Optimisation}\label{s:mathematical-optimisation}
Subclasses of the general \ROP{} are made by restricting the properties of \Criteria, \Parameter s, \Depend{} relations, and the \DecisionRule. 

Restrictions can be intended to narrow down the informal reading of the \Criteria, \Parameter s, and \Depend{} relations, and thereby the informal interpretation of an \ROP{} as of a problem statement meaningful from the perspective of the \ZJ{} view of \RE. 

For example, the \ZJ{} view distinguishes between requirements, domain knowledge, and specification. To capture this and define an \ROP{} inspired by the \ZJ{} view, the following would need to be done. All \Criteria{} and \Parameter s need to be binary variables. The \Criterion{} class should have two subclasses, requirement and domain knowledge. Specification should be the only subclass of the \Parameter{} class.

Another kind of restrictions, independent from what one thinks \RE{} is about, are on types of \Criterion{} and \Parameter{} variables, and the mathematical properties of \Depend{} relations. They are interesting, because the general \ROP{} is a subclass of the general mathematical optimisation problem, defined as follows.

\begin{definition}\label{d:general-optimisation=problem}
\xb{Optimisation Problem \cite{boyd2004convex}:} A mathematical optimisation problem, or just an Optimisation Problem, has the following form:

\begin{eqnarray*}
& & \zi{Minimise }\; f_{0}(x) \\
& & \zi{subject to }\; f_{i}(x) \leq b_{i}, \zi{ for } i = 1, \ldots, n
\end{eqnarray*}

where the vector $x = (x_{1}, \ldots, x_{n})$ is called the optimisation variable of the problem, the function $f_{0}(x)$ is called the objective function, the functions $f_{1}(x) \leq b_{1}, \ldots, f_{m}(x) \leq b_{m}$, are the constraint functions, and the constants $b_{1}, \ldots, b_{m}$ are the limits, or bounds, for the constraints.
\end{definition}

\begin{definition}\label{d:general-optimal-solution}
\xb{Optimal Solution :} The Optimal Solution to an Optimisation Problem is the vector $x^{\ast}$ if it has the smallest objective value among all vectors that satisfy the constraints: that is, for any $x^{\prime}$ with $f_{1}(x^{\prime}) \leq b_{1}, \ldots, f_{m}(x^{\prime}) \leq b_{m}$, we have $f_{0}(x^{\prime}) \geq f_{0}(x^{\ast})$.
\end{definition}

\ROP{} differs from the above in (i) having a different terminology, and (ii) equality in constraints. These differences still make the \ROP{} a subclass of the General Optimisation Problem above. We can therefore reuse resolution techniques from mathematical optimisation \cite{bradley1977applied, boyd2004convex, krentel1988complexity, papadimitriou1991optimization, yannakakis1991expressing, junger201050}. For example:

\begin{enumerate}
    \item{Depending on \Criteria{} and \Parameter{} variable types, we can have the following \ROP{} subclasses:
    \begin{enumerate}
        \item{Binary \ROP, where all \Criteria{} and \Parameter{} variables have binary value,}
        \item{Integer \ROP, where all variables take an integer value. We can have this if we allow more than two levels of satisfaction of \Criteria, and more than two configuration values for \Parameter s,}
        \item{Continuous \ROP, if all Criteria and Parameters take real numbers as values,}
        \item{Mixed-integer \ROP, if there are binary, integer, and continuous variables in the \DecisionSet.}
    \end{enumerate}
    }
    \item{\Depend{} relation properties give another classification dimension, where we can distinguish:
    \begin{enumerate}
        \item{Linear \Depend{} relations, where every relation is a linear function,}
        \item{Nonlinear \Depend{} relations, where every relation is an arbitrary nonlinear function,}
        \item{General \Depend{} relations, where some relations are linear, others nonlinear.}
    \end{enumerate}
    }
\end{enumerate}

\section{\ROP{} and Decision Analysis}\label{s:decision-analysis}
``Decision analysis is a logical procedure for the balancing of the factors that influence a decision. The procedure incorporates uncertainties, values and preferences in a basic structure that models the decision.'' \cite{howard1966decision}

The decision analysis procedure involves four steps \cite{keeney1982decision, keeney1993decisions}. In step one, the aims are to specify the objectives to achieve by taking the decision, generate alternatives to choose from in order to achieve the objectives, and identify attributes used to measure the degree to which the objectives are achieved. Step two measures the uncertainty of the consequences of alternatives; the aim is to quantify uncertainty with a probability distribution of attribute values, with one probability distribution per alternative. Step three captures the relative desirability of value assignments to attributes. This gives a utility function, as a function over attributes. Step four ranks alternatives by their expected utility, by following the rule that the higher its expected utility, the more desirable the alternative. 

The problem in decision analysis can be formulated as an Optimisation Problem, as follows.

\begin{definition}\label{d:daop}
\xb{Decision Analysis Optimisation Problem (DAOP)} has the following form:

\begin{eqnarray*}
& & \zi{Maximise }\; \sum_{i=1}^{i=n}(p(x_{j,i}) * U(x_{j,i})) \\
& & \zi{subject to }\; \sum_{i=1}^{i=n}p(x_{j,i}) = 1, \zi{ for } i = 1, \ldots, m
\end{eqnarray*}

where $j$ denotes an alternative among $m$ alternatives, $i$ an attribute among $n$ attributes, $x_{j,i}$ the value of the attribute $i$ when alternative $j$ is chosen, $p(x_{j,i})$ the probability of observing $x_{j,i}$ for $i$ when choosing alternative $j$, and $U(x_{j,i})$ the utility of observing $x_{j,i}$ for $i$ when choosing alternative $j$. 
\end{definition}

\begin{definition}\label{d:daop-solution}
\xb{Optimal Solution to DAOP} is an alternative $j^{\ast}$ such that for any other alternative $j^{\prime}$, it is true that $\sum_{i=1}^{i=n}( p(x_{j^{\prime},i}) * U(x_{j^{\prime},i}) ) \leq \sum_{i=1}^{i=n}( p(x_{j^{\ast},i}) * U(x_{j^{\ast},i}) )$.
\end{definition}

The relationship between \ROP{} and DAOP is that there are subclasses of \ROP{} which are also subclasses of DAOP.

There are many such subclasses, and they depend on how \ROP{} concepts are mapped to DAOP concepts. Here are the steps to follow, to make one such \ROP{} subclass:
\begin{enumerate}
    \item{Let every alternative $j = 1, \ldots, m$ in the DAOP be a point in the \SolutionSpace.} 
    \item{Let each attribute $i = 1, \ldots, n$ that is used to evaluate alternatives, be a \Criterion{} in the \SolutionSpace.}
    \item{Add one \Criterion{} to the \SolutionSpace, and call it utility. Its values are utility values, interpreted informally in the same way utility values are in decision analysis. Note that the \SolutionSpace{} now has $n+1$ dimensions.}
    \item{Add \Depend{} relations, which map values of all non-utility \Criteria, or combinations of these \Criteria, to values of the utility \Criterion. This is the same as saying, using terms common to \RE, that different levels of satisfaction of requirements and domain assumptions that these non-utility \Criteria{} amount to, result in different levels of utility. It is necessary to add these \Depend{} relations, because in decision analysis, utility is the aggregate measure of the attributes used to evaluate alternatives, and these relations will reflect this. If such Relations were absent, it will look like the \DecisionRule{} disregards all \Criteria{} other than utility, or disregards all those Criteria whose values map to no utility value.}
    \item{Constraints $\sum_{i=1}^{i=n}p(x_{j,i}) = 1$ for $j = 1, \ldots, m$ from DAOP should be carried over as \Depend{} relations to the \ROP.}
    \item{The \DecisionRule{} of the \ROP{} should be equal to the \Depend{} relation that corresponds to the objective function in the DAOP.}
\end{enumerate}

The resulting \ROP{} subclass is a subclass of DAOP, in the sense that the only differences between DAOP and the \ROP{} subclass are those of terminology and in the number of \Depend{} relations, since it is necessary to add \Depend{} relations that map non-utility \Criteria{} values to the utility \Criterion{} values.

\section{\ROP{} and Expected Utility Theory}\label{s:expected-utility-theory}
``Expected utility models are concerned with choices among risky prospects whose outcomes may be either single or multidimensional. If we denote these various (say $n$) outcome vectors by $x_{i}$ and denote the $n$ associated probabilities by $p_{i}$ such that the sum over $i = 1$ to $i = n$ of $p_{i}$ equals 1, we then generally define an Expected Utility (EU) model as one which predicts or prescribes that people maximize the sum over $i = 1$ to $i = n$ of $F(p_{i}) * U(x_{i})$. [...] Within this general EU model different variants exist depending on (1) how utility is measured, (2) what kind of probability transformations F(.) are allowed, and (3) how the outcomes $x_{i}$ are measured." [Schoemaker:1982]

The DAOP introduced in the previous section is equivalent to the optimisation problem central to EU, so that the same remark as for DAOP applies: there is a subclass of SSOP which is also a subclass of the EU problem, and we can proceed to define that subclass in the way as for DAOP.

\section{Conclusions}
This paper argued that the \RPASfull{} is different from the \ZJRP, and that it is not a subclass of the \ZJRP. It then proposed a general definition of the \RPAS. Finally, the paper related \RPAS{} to mathematical optimisation in general, to decision analysis in management science, and to expected utility theory in economics.

\xb{Acknowledgements.} The first author is funded by the Fonds de la Recherche Scientifique -- FNRS (Brussels, Belgium). This work was supported in part by ERC advanced grant 267856, titled ``Lucretius: Foundations for Software Evolution''.

\bibliographystyle{plain}
\bibliography{references}

\end{document}